\RequirePackage[2020-02-02]{latexrelease}
\documentclass[twocolumn]{revtex4}

\usepackage{color}
\newcommand{\REV}[1]{{\color{black} #1}}
\newcommand{\ket}[1]{\left| #1 \right\rangle}

\usepackage{graphicx}
\usepackage{amsmath}
\usepackage{amssymb}

\newcommand{\AppOBE}{Appendix~\ref{sec:OBE}}
\newcommand{\AppEtalon}{Appendix~\ref{sec:Etalon}}
\newcommand{\AppOD}{Appendix~\ref{sec:OD}}
\newcommand{\AppUACF}{Appendix~\ref{sec:UACF}}

\newcommand{\CiteQC}{QC1, QC2, QC3, QC4}
\newcommand{\CiteQM}{QM1, QM2, QM3, QM4, QM5}
\newcommand{\CiteQWC}{QWC1, QWC2, QWC3, QWC4, QWC5}
\newcommand{\CiteDtColdAtom}{
PRL96(2006),
J.Opt.Soc.Am.B24(2007),
PRA93(2016),
OE25(2017)}
\newcommand{\CiteDtHotAtom}{
PRA82(2010),
OE19(2011),
OE20(2012),
Optica2(2015),
OE28(2020),
Adv.Q7(2024),
PRA21(2024)}
\newcommand{\RT}[1]{``#1,''}

\begin{document}
\title{Temporally-long C-band heralded single photons generated from hot atoms}

\author{
Pei-Yu Tu,$^1$ 
Chia-Yu Hsu,$^1$
Wei-Kai Huang,$^1$
Tse-Yu Lin,$^1$ 
Chih-Sung Chuu,$^{1,2}$
Ite A. Yu$^{1,2,}$}\email{yu@phys.nthu.edu.tw}

\affiliation{
$^1$Department of Physics, National Tsing Hua University, Hsinchu 30013, Taiwan \\
$^2$Center for Quantum Science and Technology, National Tsing Hua University, Hsinchu 30013, Taiwan
}

\begin{abstract}
C-band photons are recognized for having the lowest loss coefficient in optical fibers, making them highly favorable for optical fiber-based communication. In this study, we systematically investigated the temporal width of C-band heralded single photons and developed a theoretical model for biphoton generation via the spontaneous four-wave mixing (SFWM) process using a diamond-type transition scheme, which has not been previously reported. Our experimental data on temporal width closely aligns with the predictions of this model. Additionally, we introduced a new concept: the atomic velocity group relating to the two-photon resonance condition and the one-photon detuning in this atomic frame. These two parameters are crucial for understanding the behavior of the biphoton source. The concept indicates that the hot-atom source behaves similarly to the cold-atom source. Guided by our theoretical model, we observed 1529-nm (C-band) heralded single photons with a temporal width of 28.3$\pm$0.6 ns, corresponding to a linewidth of 11.0$\pm$0.2 MHz. For comparison, the ultimate linewidth limit is 6.1 MHz, determined by the natural linewidth of the atoms. Among all atom-based sources of 1300 to 1550 nm heralded single photons utilizing either cold or hot atoms, the temporal width achieved in this work represents the first instance of a width exceeding 10 ns, making it (or its linewidth) the longest (or narrowest) record to date. This work significantly enhances our understanding of diamond-type or cascade-type SFWM biphoton generation and marks an important milestone in achieving greater temporal width in atom-based sources of C-band heralded single photons.
\end{abstract}

\maketitle

\newcommand{\Table}{
	\begin{table}[t]
	\caption{Temporal FWHMs of atom-based SFWM biphoton sources using the diamond-type or cascade-type transition schemes.}
	{\centering
	\begin{tabular}{c c c c c}
	\hline\hline
	\parbox{14mm}{\vspace*{2mm}Reference\\Number \vspace*{2mm}} & 
	\parbox{14mm}{Transition\\Scheme} &
	\parbox{14mm}{Medium$^\ast$} &
	\parbox{20mm}{Wavelengths\\(nm)} & 
	\parbox{17mm}{Temporal\\FWHM~(ns)$^\dagger$} \\
 	\hline
	This work & 
		Diamond & RHA & 1529 \& 780 & 28.3$\pm$0.6 \\
	\onlinecite{OE25(2017)} & 
		Diamond & LCA & 1476 \& 795 & 7 \\
	\onlinecite{PRL96(2006)} & 
		Cascade & LCA & 1529 \& 780 & 4.6 \\
	\onlinecite{PRA93(2016)} & 
		Diamond & LCA & 1476 \& 795 & 3.5 \\
	\onlinecite{OE28(2020)} & 
		Diamond & RHA & 1529 \& 780 & 1 \\
	\onlinecite{Optica2(2015)} & 
		Diamond & RHA & 1529 \& 780 & 1 \\
	\onlinecite{OE20(2012)} & 
		Cascade & RHA & 1529 \& 780 & 1 \\
	\onlinecite{OE19(2011)} & 
		Diamond & RHA & 1367 \& 780 & 1 \\
	\onlinecite{PRA82(2010)} & 
		Diamond & RHA & 1367 \& 780 & 1 \\
	\onlinecite{Adv.Q7(2024)} & 
		Cascade & RHA & 1529 \& 780 & 0.6 \\
	\onlinecite{PRA21(2024)} & 
		Diamond & RHA & 1324 \& 795 & 0.5 \\
	\hline\hline	
	\end{tabular} 
	\\
	\vspace*{-0.5\baselineskip}
	\hspace*{3.0pt}
	\parbox{85mm}{
	\begin{flushleft}
	\hspace*{-4.0pt}$^{\ast}$RHA: room-temperature or hot atoms. 
		LCA: laser-cooled atoms.\\
		\hspace*{-4.0pt}$^{\dagger}$In each reference, we list the best temporal width 
		of the biphoton wave packet that violates the Cauchy-Schwarz inequality 
		for classical light.
	\end{flushleft}
	}
	}
	\label{table:One}
	\end{table}
}

\newcommand{\FigOne}{
	\begin{figure}[b]
	\includegraphics[width=\columnwidth]{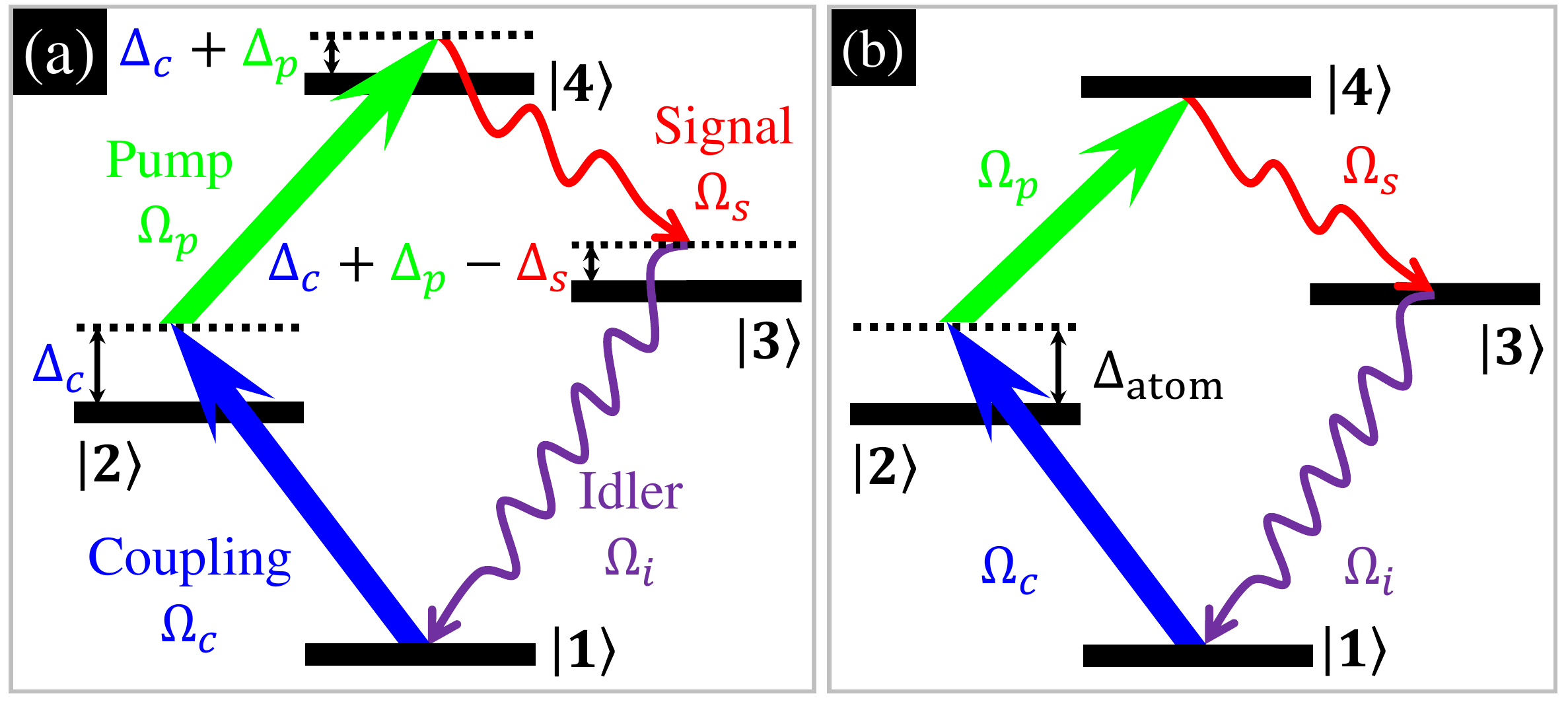}
	\caption{
    (a) The diamond-type transition scheme or cascade-type scheme, where $|2\rangle$ and $|3\rangle$ are the same state. In the SFWM process, the coupling and pump laser fields, along with the signal and idler single photons form the four-photon resonance, where $\Delta_c$, $\Delta_p$, $\Delta_s$, and $\Delta_i$ (= $\Delta_c+\Delta_p-\Delta_s$) represent their one-photon detunings, and $\Omega_c$, $\Omega_p$, $\Omega_s$, and $\Omega_i$ are the Rabi frequencies. The actual energy levels of the atoms in the experiment are specified in Sec.~\ref{sec:setup}. (b) The transition scheme viewed in the atom frame, where the Doppler shift of the atoms causes the coupling and pump fields to satisfy the two-photon resonance. $\Delta_{\rm atom}$ represents the one-photon detuning of the coupling-pump transition. The signal (or idler) photon decays with a frequency near the resonance frequency of the transition $|4\rangle$$\rightarrow$$|3\rangle$ (or $|3\rangle$$\rightarrow$$|1\rangle$). 
	}
	\label{fig:Transition}
	\end{figure}
}
\newcommand{\FigTwo}{
	\begin{figure}[t]
	\includegraphics[width=\columnwidth]{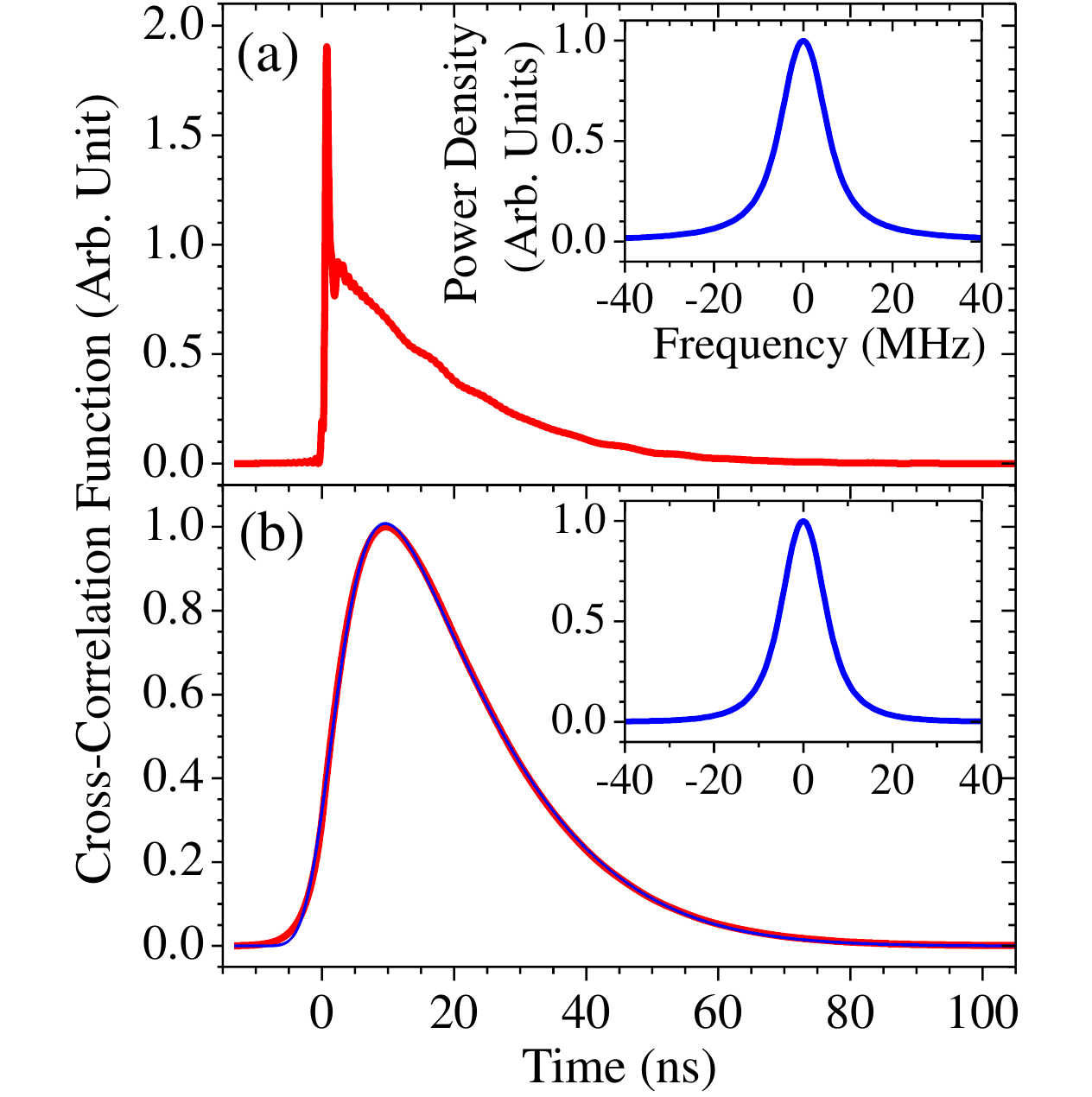}
	\caption{
    Theoretical predictions (red lines) of the representative two-photon correlation functions, or biphoton wave packets (a) before and (b) after passing through the etalons, whose characteristics are illustrated in \AppEtalon. Since the signal photon appears before the idler photon, we calculated the generation probability of the idler photon right after the atomic vapor cell upon the appearance of the signal photon. Each inset shows the spectrum that gives the wave packet in the main plot. In the calculation, $(\alpha, \Omega_c, \Omega_p, \gamma)$ = $(420, 17\Gamma, 78\Gamma, 0.4\Gamma)$ and $(\omega_{D0}, \Delta_{\rm atom})$ = $(0, 380\Gamma)$ or 2$\pi$$\times$(0, 2.28~GHz). The spectral FWHMs are (a) 12.0 and (b) 11.4~MHz. The blue line in the main plot of (b) represents the best fit, with an FWHM of 25.9~ns.
    }
	\label{fig:Theory_WP}
	\end{figure}
}
\newcommand{\FigThree}{
	\begin{figure}[t]
	\includegraphics[width=\columnwidth]{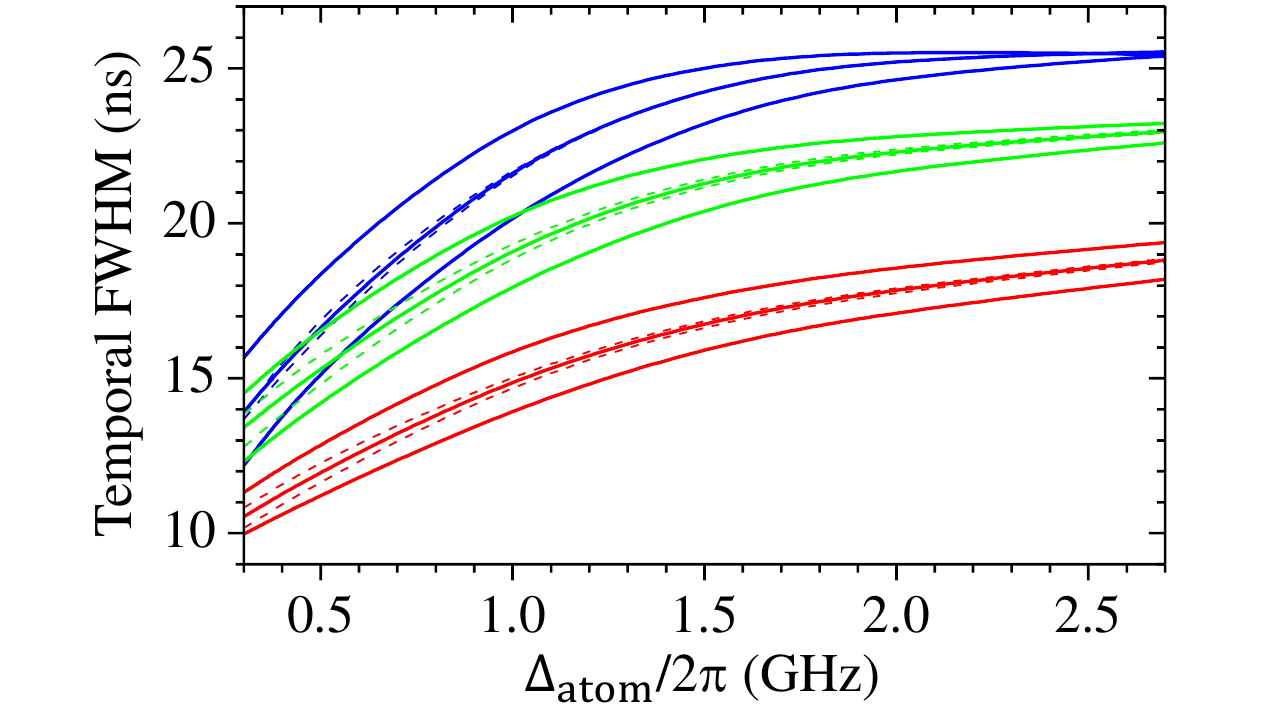}
	\caption{
    Theoretical predictions of the temporal FWHM as a function of $\Delta_{\rm{atom}}$ for different values of $\omega_{D0}$.  Blue, green, and red lines correspond to $\omega_{D0}/2\pi$ = 0, 0.24, and 0.48~GHz, respectively. For all cases, we set $\alpha$ = 400 and $\gamma$ = 0.4$\Gamma$. We set $\Omega_c\times\Omega_p$ = 1300$\Gamma^2$ for the central solid and nearby dashed lines and increased (decreased) the value by 20\% for the lower (upper) solid lines. We also set $\Omega_p/\Omega_c$ = 4.6 for the solid lines and increased (decreased) the value by 20\% for the upper (lower) dashed lines. The transition dipole matrix element, Clebsch-Gordan coefficients, and ratio of pump to coupling powers determine $\Omega_p/\Omega_c$ = 4.6.
    }
	\label{fig:Theory_Delta}
	\end{figure}
}
\newcommand{\FigFour}{
	\begin{figure}[t]
	\includegraphics[width=\columnwidth]{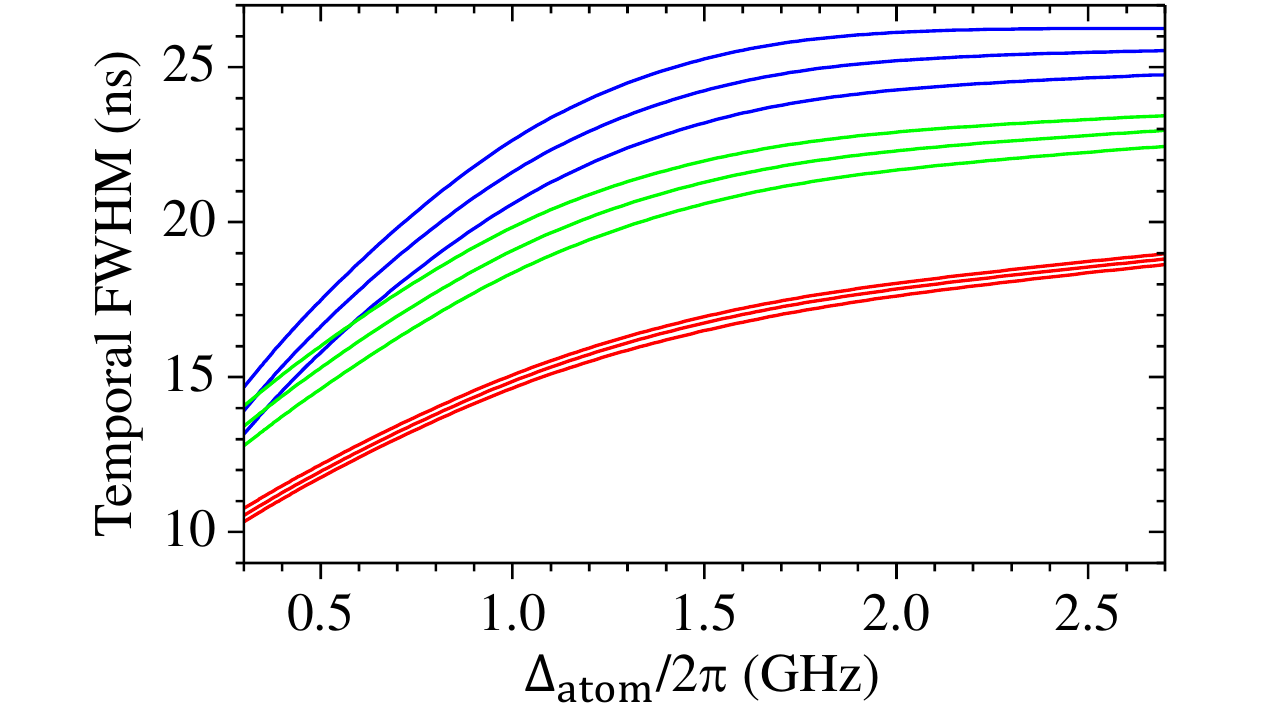}
	\caption{
    Effect of the optical depth $\alpha$ on the temporal FWHM. Blue, green, and red lines represent $\omega_{D0}/2\pi$ = 0, 0.24, and 0.48~GHz, respectively. For all lines, $\gamma$ = 0.4$\Gamma$, $\Omega_c \times \Omega_p$ = 1300$\Gamma^2$, and $\Omega_p/\Omega_c$ = 4.6. We set $\alpha$ = \REV{450}, 400, and \REV{350} for the upper, central, and lower lines of the same color.
    }
	\label{fig:Theory_OD}
	\end{figure}
}
\newcommand{\FigFive}{
	\begin{figure}[t]
	\includegraphics[width=\columnwidth]{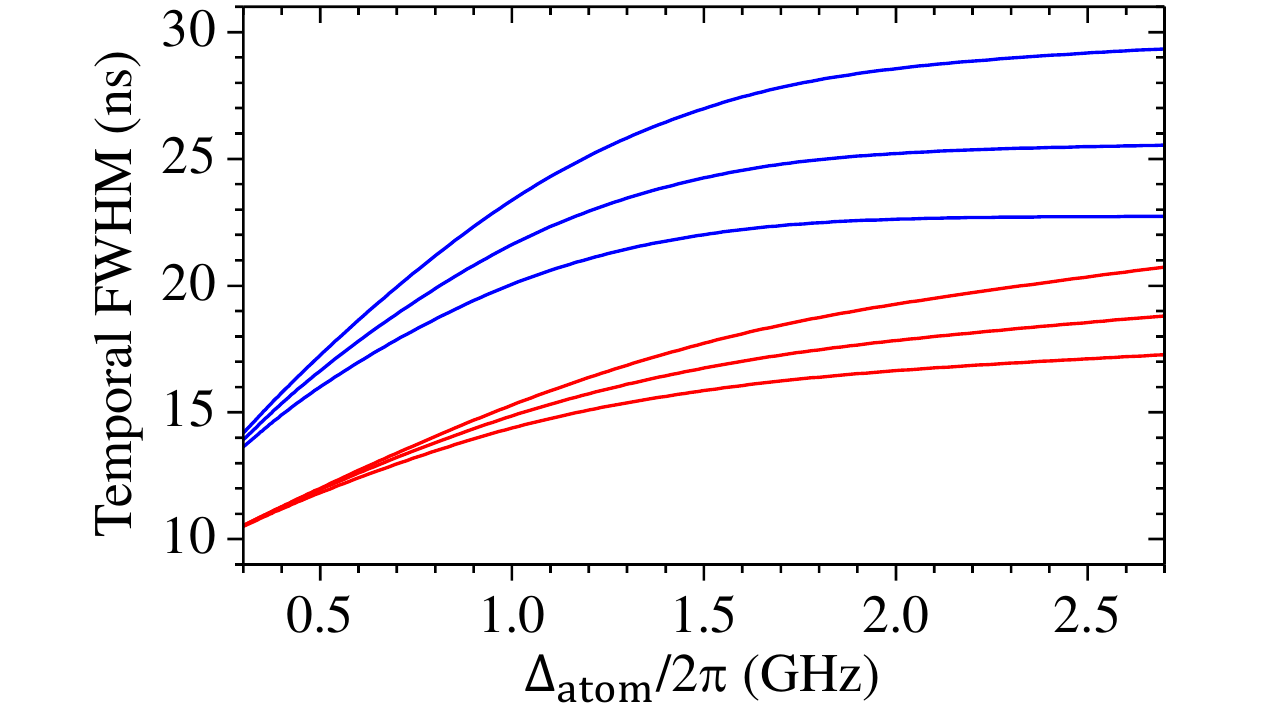}
	\caption{
    Effect of the decoherence rate $\gamma$ on the temporal FWHM. Blue and red lines represent $\omega_{D0}/2\pi$ = 0 and 0.48~GHz, respectively.  For all lines, $\alpha$ = 400, $\Omega_c$$\times$$\Omega_p$ = 1300$\Gamma^2$, and $\Omega_p/\Omega_c$ = 4.6. We set $\gamma$ = 0.2, 0.4, and 0.6 for the upper, central, and lower lines of the same color, respectively.
    }
	\label{fig:Theory_gamma}
	\end{figure}
}
\newcommand{\FigSix}{
	\begin{figure}[t]
	\includegraphics[width=\columnwidth]{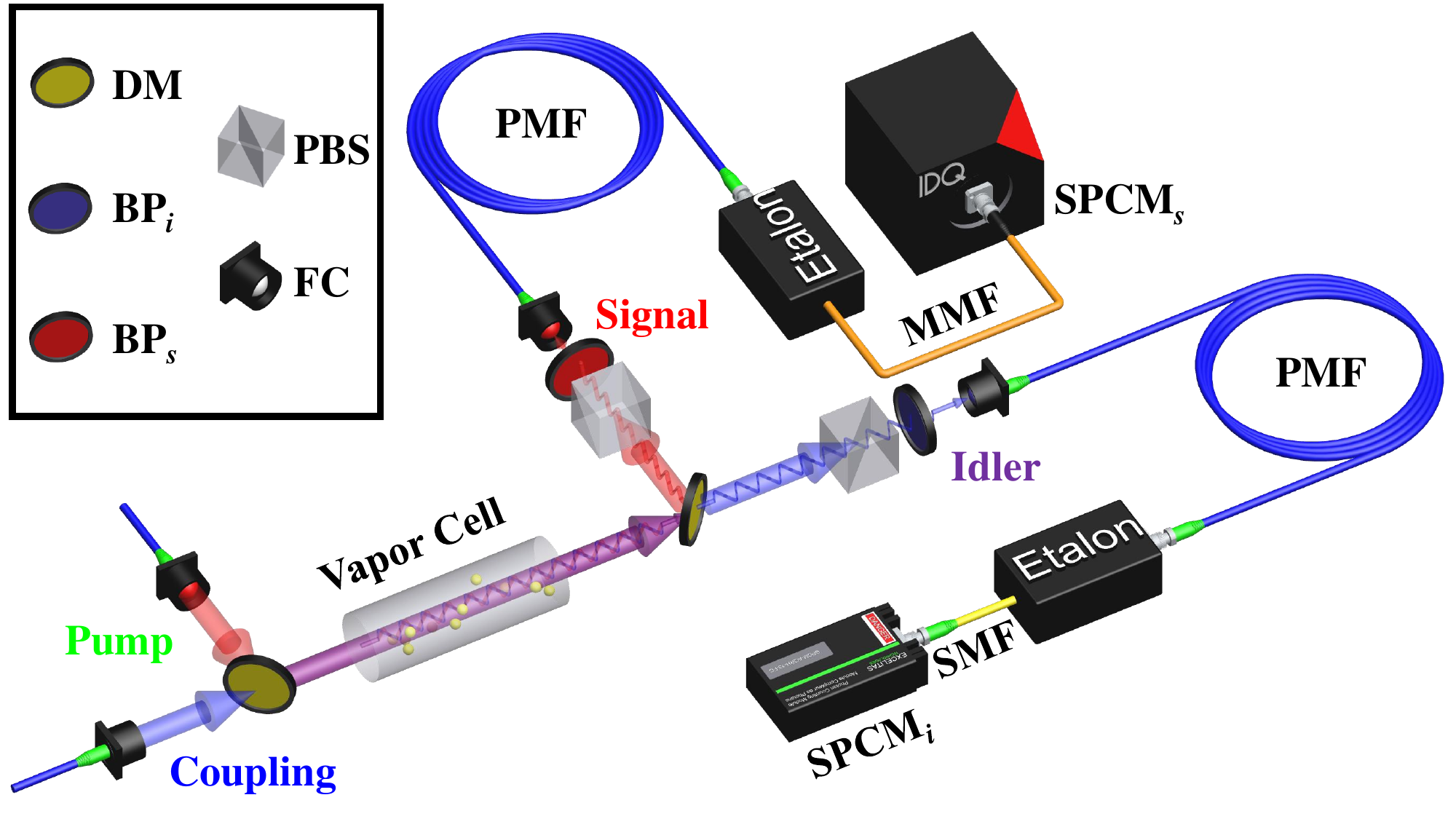}
	\caption{
    Schema of the experimental setup: DM: dichroic mirror, BP$_i$: three bandpass filters for the idler photons, BP$_s$: three bandpass filters for the signal photons, PBS: polarization beam splitter, FC: optical fiber collimation lens, PMF, SMF, and MMF: polarization-maintained, single-mode, and multimode optical fibers, respectively; SPCM$_s$ and SPCM$_i$: signal and idler single-photon counting modules.
	}
	\label{fig:ExpSetup}
	\end{figure}
}
\newcommand{\FigSeven}{
	\begin{figure}[t]
	\includegraphics[width=\columnwidth]{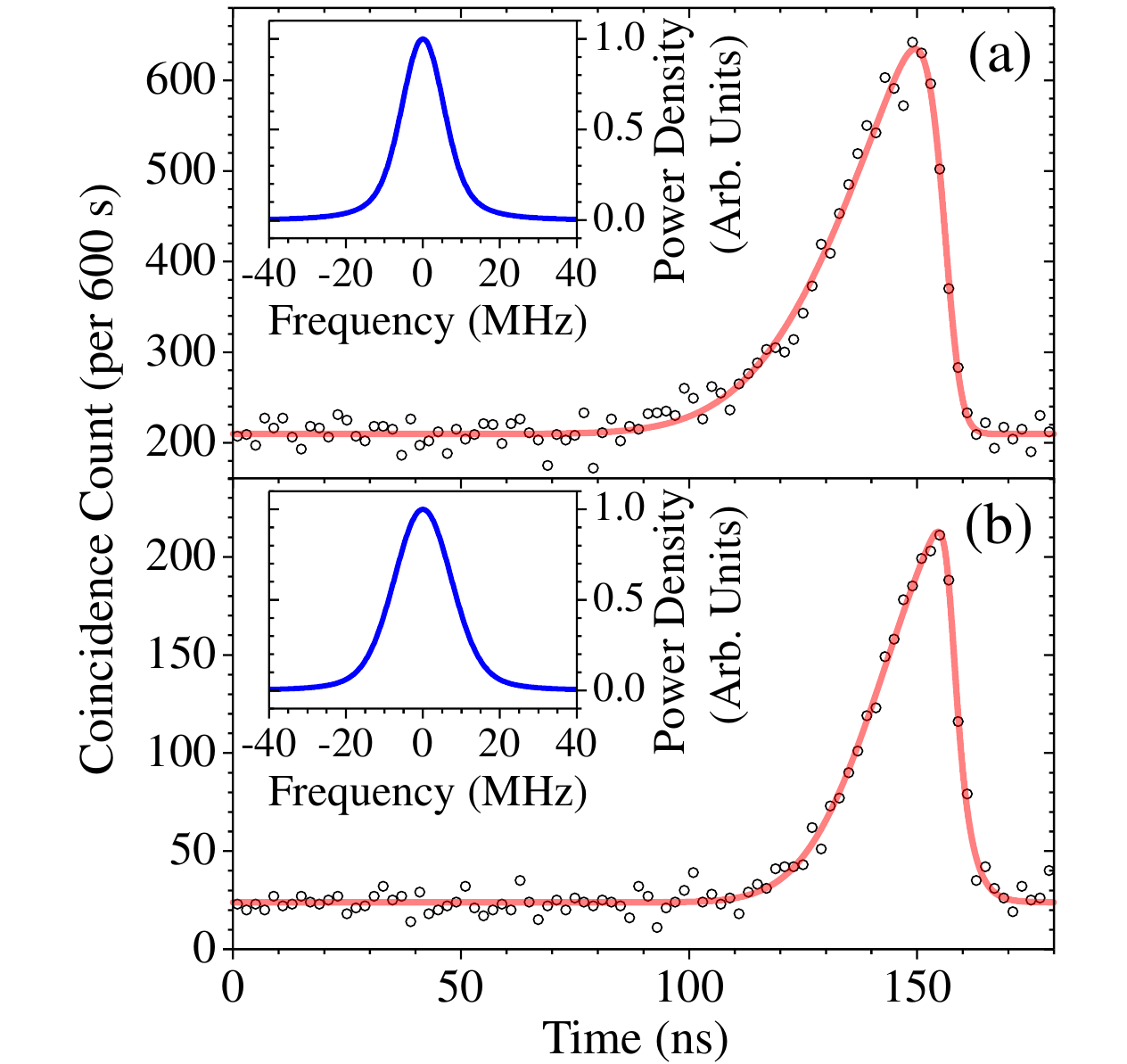}
	\caption{
    Representative data of the biphoton wave packet, or the coincidence count as a function of the delay time. Circles represent the experimental data, and red lines are the best fits using phenomenological functions. Insets show the frequency spectra corresponding to the red lines. The coupling and pump powers were 30 and 60 mW, respectively, and the OD was approximately 420 during the measurements. (a) For this case, we set $\omega_{D0} =$ 0 and $\Delta_{\rm atom}/2\pi =$ 2.28~GHz. The best fit and its spectrum indicate that the biphotons had a temporal FWHM of 25.7$\pm$0.4~ns, a spectral FWHM of 12.5$\pm$1.2~MHz, and a SBR of 1.99$\pm$0.06. (b) For this case, we set $\omega_{D0}/2\pi =$ 0.48~GHz and $\Delta_{\rm atom}/2\pi =$ 1.48~GHz. The best fit and its spectrum indicate that the biphotons had a temporal FWHM of 17.5$\pm$2.3 ns, a spectral FWHM of 16.3$\pm$1.5 MHz, and a SBR of 8.5$\pm$0.6. 
	}
	\label{fig:Data_WP}
	\end{figure}
}
\newcommand{\FigEight}{
	\begin{figure}[t]
	\includegraphics[width=\columnwidth]{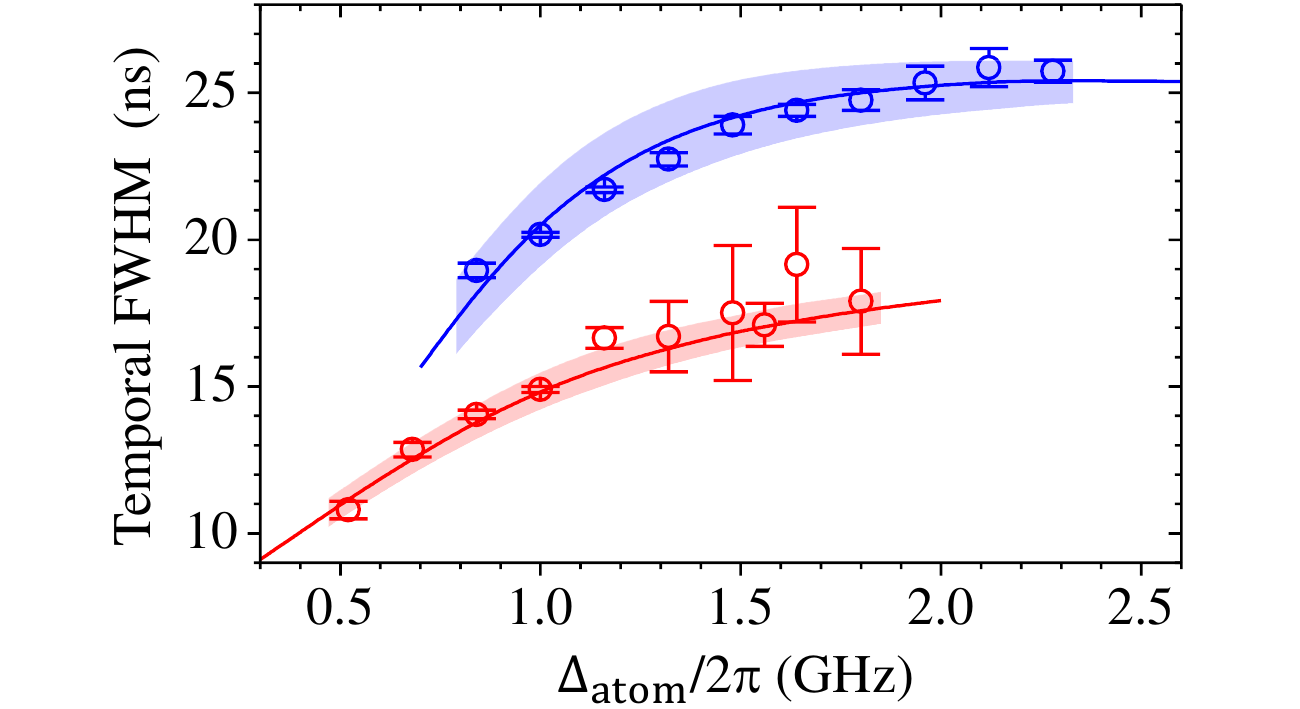}
	\caption{
    The temporal FWHM of biphoton wave packet as a function of $\Delta_{\rm{atom}}$. Red and blue circles represent the experimental data measured at $\omega_{D0}/2\pi =$ 0 and 0.48~GHz, respectively, Error bars represent the mean deviations of the corresponding data. Solid lines are the theoretical predictions calculated from Eq.~(\ref{eq:GtwoEtalon}). In the calculation, $\Omega_c = 17.1$$\Gamma$, $\Omega_p =$ 4.6$\Omega_c$, $\gamma$ = 0.43$\Gamma$, and the values of $\alpha$ (OD) followed the experimental results presented in \AppOD. Shaded areas represent the uncertainties in $\Omega_c$$\times$$\Omega_p$ and $\alpha$, where we decreased (increased) $\Omega_c$$\times$$\Omega_p$ by 10\% and enhanced (reduced) $\alpha$ by 10\% to calculate their upper (lower) bounds.
    }
	\label{fig:Data_tFWHM}
	\end{figure}
}
\newcommand{\FigNine}{
	\begin{figure}[t]
	\includegraphics[width=\columnwidth]{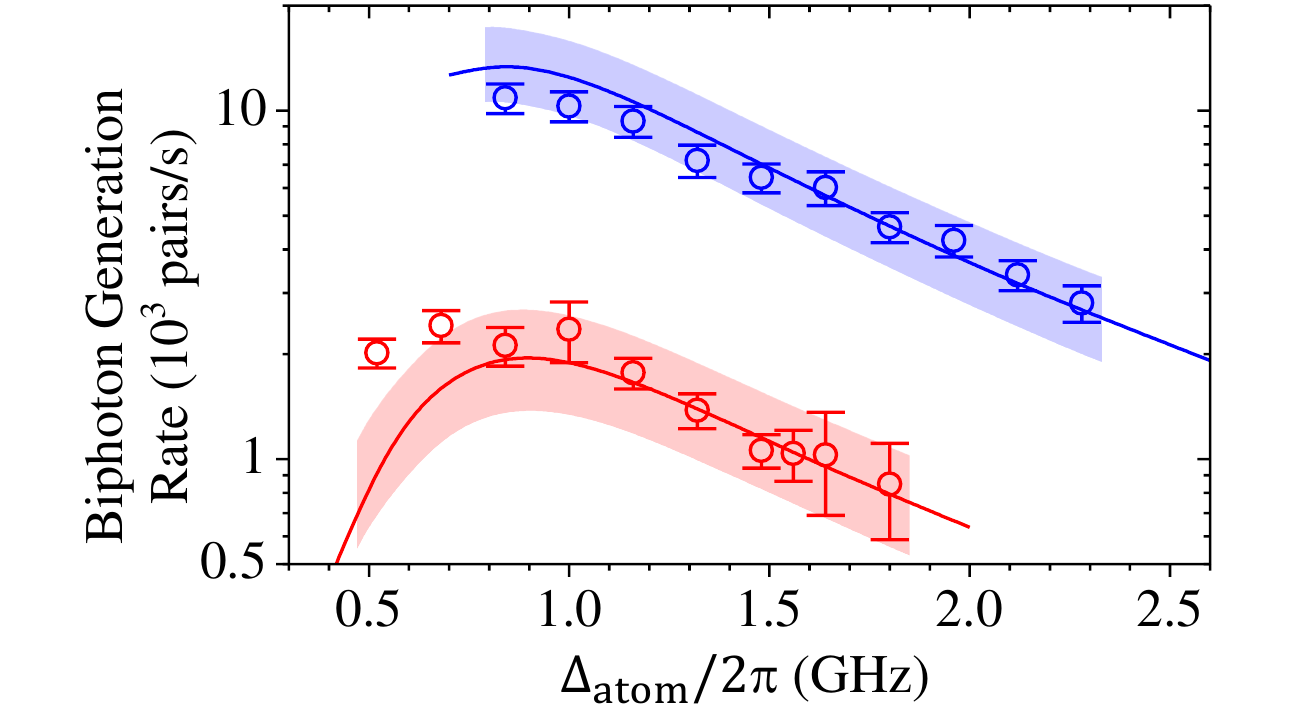}
	\caption{
    Biphoton generation rate as a function of $\Delta_{\rm{atom}}$. The rate refer to the biphotons collected inside the polarization-maintained optical fibers. The legends and the calculation parameters of theoretical predictions are the same as those specified in the caption of Fig.~\ref{fig:Data_tFWHM}, except that we increased (decreased) $\Omega_c$$\times$$\Omega_p$ by 10\% and enhanced (reduced) $\alpha$ by 10\% to calculate the upper (lower) bounds of shaded areas.
    }
	\label{fig:Data_GR}
	\end{figure}
}
\newcommand{\FigTen}{
	\begin{figure}[t]
	\includegraphics[width=\columnwidth]{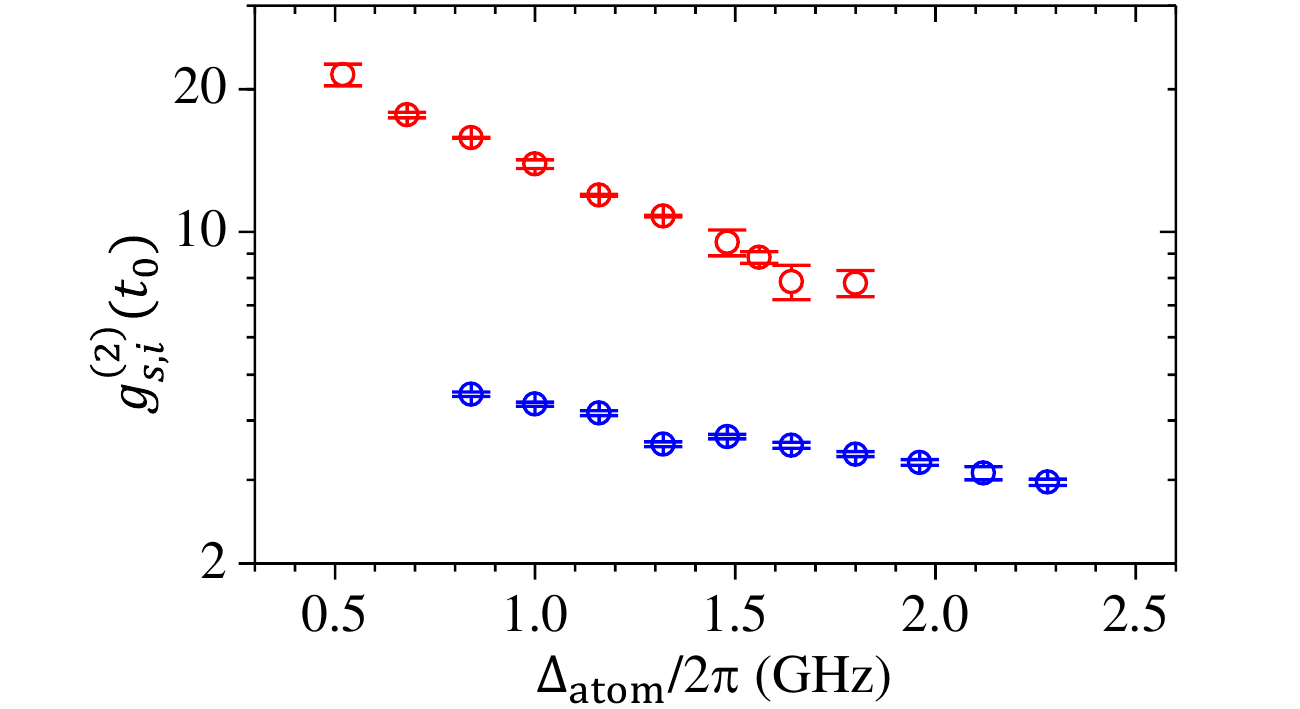}
	\caption{
    Experimental data for the peak of the cross-correlation function, $g^{(2)}_{s,i}(t_0)$, as a function of $\Delta_{\rm{atom}}$. The red and blue circles represent the experimental data measured at $\omega_{D0}/2\pi =$ 0 and 0.48~GHz, respectively. The error bars represent the mean deviations of the corresponding data.
    }
	\label{fig:Data_CCF}
	\end{figure}
}
\newcommand{\FigEleven}{
	\begin{figure}[t]
	\includegraphics[width=\columnwidth]{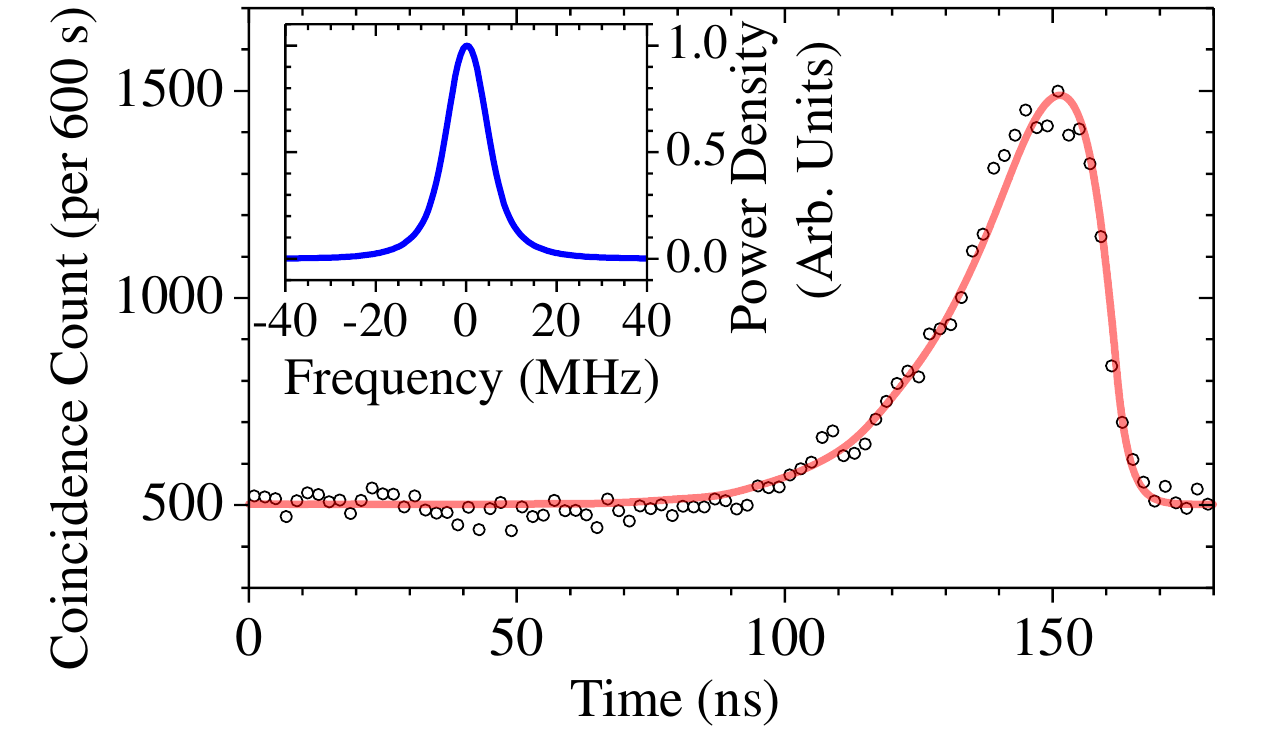}
	\caption{The biphoton wave packet measured after the alignment optimization (see text), where $\omega_{D0} =$ 0 and $\Delta_{\rm atom}/2\pi =$ 2.12~GHz. Circles represent the experimental data; the red line represents the theoretical prediction. We calculated the prediction from the frequency spectrum shown in the inset using Eqs.~(\ref{eq:GtwoEtalon}) and (\ref{eq:BiphotonSpectrum}). In the calculation, the values of OD, $\Omega_c$, and $\Omega_p$ are the same as those in Fig.~\ref{fig:Data_tFWHM}, and the value of $\gamma$ was varied. To best match the prediction to the data, $\gamma$ = (0.27$\pm$0.03)$\Gamma$. The biphotons had a temporal FWHM of 28.3$\pm$0.6~ns, a spectral FWHM of 11.0$\pm$0.2~MHz, an SBR of 2.00$\pm$0.05, and a generation rate of (7.6$\pm$0.8)$\times$$10^3$ pairs/s.
    }
	\label{fig:LongestWP}
	\end{figure}
}

\newcommand{\FigTwelve}{
	\begin{figure}[t]
	\includegraphics[width=\columnwidth]{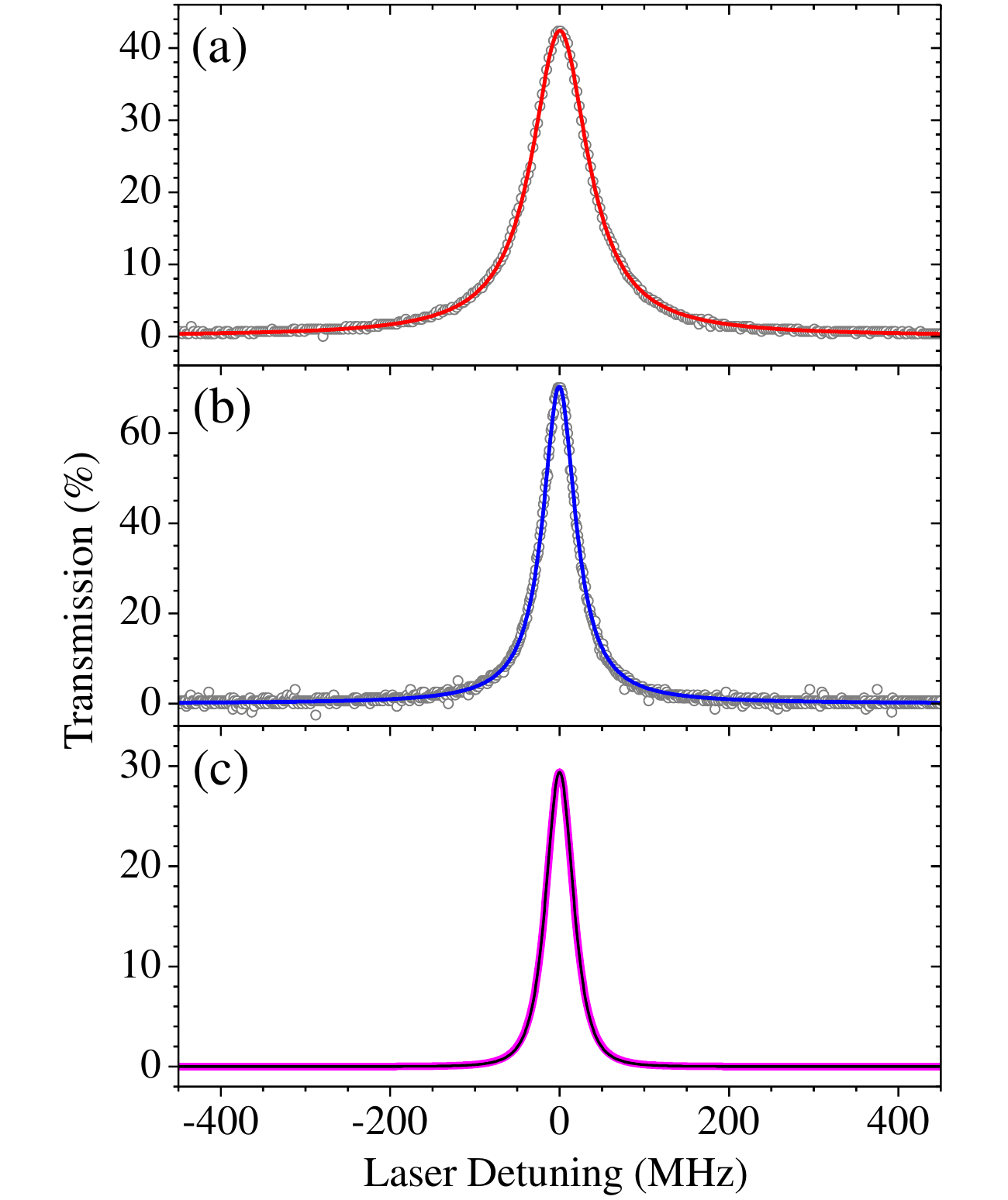}
	\caption{
    Power transmission spectra of (a) the 1529-nm etalon, (b) the 780-nm etalon, and (c) the two etalons combined. Circles are the experimental data. Red and blue lines represent the best fits of Lorentzian functions. The magenta line is the product of the red and blue lines, while the black line best fit of the magenta line using a squared Lorentzian function. The best fits from (a) to (c) give the peak transmissions of 42\%, 70\%, and 29\% with the FWHMs of 80, 47, and 38~MHz, respectively.
    }
	\label{fig:Etalon}
	\end{figure}
}
\newcommand{\FigThirteen}{
	\begin{figure}[t]
	\includegraphics[width=\columnwidth]{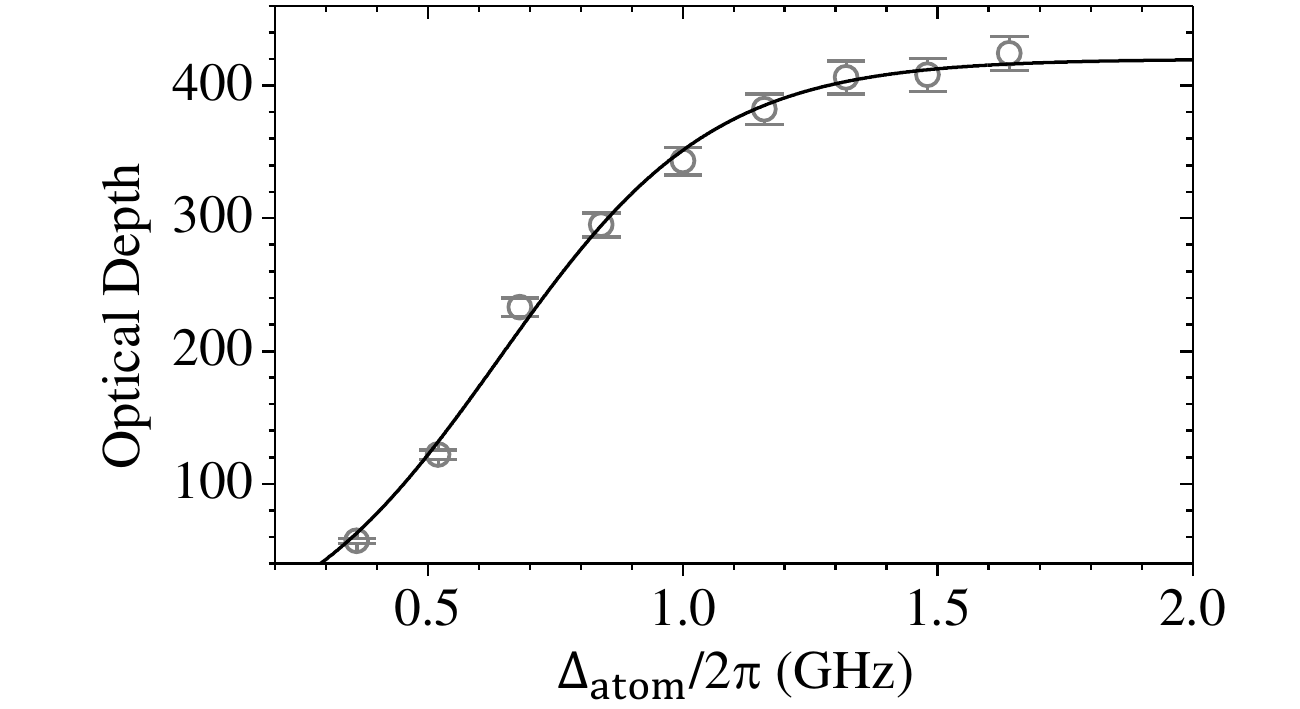}
	\caption{
    The optical depth as a function of $\Delta_{\rm{atom}}$ at $\omega_{D0} =$ 0. Circles represent the experimental data, and the solid line represents the phenomenological best fit of a hyperbolic tangent function. 
    }
	\label{fig:OD}
	\end{figure}
}
\newcommand{\FigFourteen}{
	\begin{figure}[t]
	\includegraphics[width=\columnwidth]{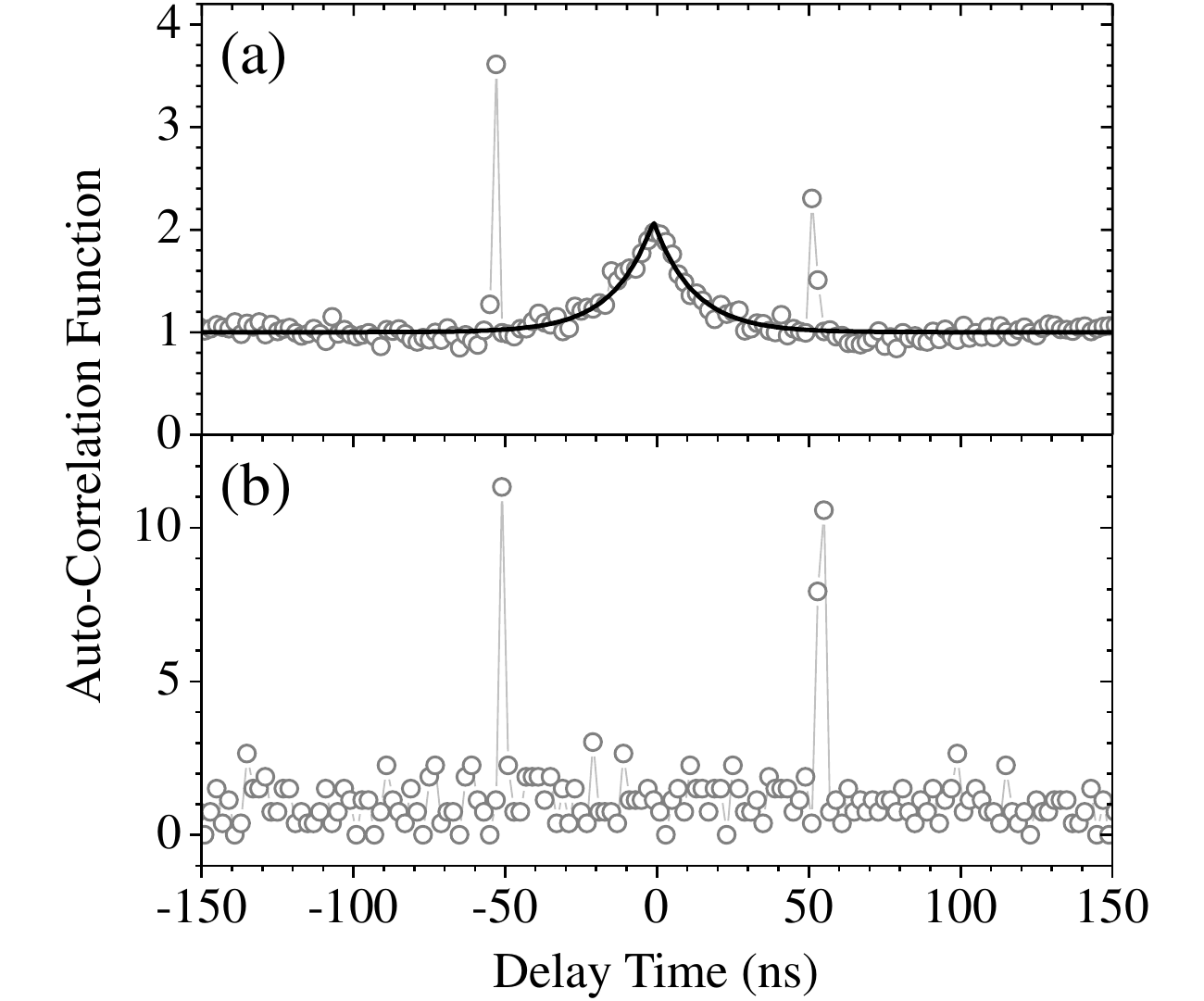}
	\caption{
    (a) The auto-correlation function of the signal photons generated from the diamond-type SFWM biphoton source and passing through the 1529-nm etalon. The experimental conditions here were the same as in Fig.~\ref{fig:Data_WP}. The circles, connected by gray lines, represent the data. The black line is the best fit to the data and has a peak value of 2.06$\pm$0.04. (b) The auto-correlation function of a weak 1529-nm laser light passing through the same etalon. The photon rate of the laser power was about 23\% of the signal photon rate in (a).
    }
	\label{fig:UACF}
	\end{figure}
}

\section{Introduction}

Biphotons are pairs of time-correlated single photons. In a biphoton pair, one photon, known as the heralding photon, starts or triggers a quantum operation, while the other, called the heralded photon, is utilized in that operation. The deterministic timing and linewidth make the heralded photons versatile qubits. The C-band heralded single photon (HSP) has garnered significant interest \cite{\CiteQC} because it experiences minimal attenuation in optical fibers compared to HSPs of other wavelength bands. Moreover, atom-based biphoton sources offer stable frequencies and tunable linewidths, making them suitable for high-storage-efficiency quantum memories \cite{\CiteQM} and high-conversion-efficiency quantum wavelength converters \cite{\CiteQWC}. Consequently, C-band HSPs generated from an atom-based biphoton source are well-suited for quantum repeater protocols and play a vital role in long-distance quantum communication via optical fibers.

The temporal width or spectral linewidth of HSPs is an important figure of merit of biphoton sources. For example, given the bandwidth of a quantum memory or quantum wavelength converter, an optimal HSP linewidth can maximize storage or conversion efficiency. This study primarily focuses on the HSP temporal width. To generate C-band HSPs from atoms \cite{\CiteDtColdAtom, \CiteDtHotAtom}, the spontaneous four-wave mixing (SFWM) process with the diamond-type or cascade-type transition scheme is commonly utilized, as depicted in Fig.~\ref{fig:Transition}(a). The two transition schemes have similar physical origins for the HSP temporal width.

\FigOne

Cold atoms, typically laser-cooled, offer a low dephasing or decoherence rate for atomic coherences in the FWM process due to negligible collision rates between atoms, relatively small Doppler broadenings, and minute relaxation processes caused by thermal motion. A low dephasing rate results in a longer HSP temporal width or, equivalently, a narrower HSP spectral linewidth compared to the HSPs generated from room-temperature or hot atoms. Several papers have reported the diamond-type or cascade-type biphoton sources using cold atoms, which will be described in the following paragraph. 

In Ref.~\onlinecite{PRL96(2006)}, the authors employed the cascade-type scheme to generate the first C-band HSPs. The temporal full width at half maximum (FWHM) of their 1529-nm HSPs was 4.6~ns. In Ref.~\onlinecite{J.Opt.Soc.Am.B24(2007)}, the same group later theoretically investigated the feasibility of using these HSPs for quantum memory. In Ref.~\onlinecite{PRA93(2016)}, the authors utilized the diamond-type scheme to generate polarization-entangled pairs of 795-nm and 1476-nm single photons. They used the 1476-nm photons to herald the storage of 795-nm photons in quantum memory and studied the cross-correlation and storage efficiency as functions of storage time. The biphoton temporal FWHM in this study was 3.5~ns. In Ref.~\onlinecite{OE25(2017)}, the authors reported hyper-entanglement in both polarization and time-frequency degrees of freedom with pairs of 795-nm and 1476-nm single photons. The HSP temporal FWHM in this study was 7~ns. So far, all diamond-type or cascade-type biphoton sources using cold atoms have temporal FWHMs of less than 10~ns.

Room-temperature or hot atom media operate in the continuous mode in contrast to the pulsing mode of cold atoms. The biphoton source from hot atoms typically exhibits a high generation or detection rate and a considerable spectral brightness. However, the Doppler effect and other thermal motion-related relaxation processes result in a significant dephasing or decoherence rate in the hot-atom source, shortening the temporal width or broadening the spectral linewidth, as shown by the diamond-type or cascade-type SFWM biphoton sources discussed in the following paragraph.

In Ref.~\onlinecite{PRA82(2010)}, the authors utilized the diamond-type scheme to generate pairs of 1367-nm and 780-nm photons. They achieved a temporal FWHM of 8.3~ns but with a peak cross-correlation of 1.7, which does not violate the Cauchy-Schwarz inequality for classical light. Their photon pairs could exhibit a nonclassical cross-correlation but with a much shorter temporal FWHM of 1~ns. The same authors further verified the nonclassicality by filtering out the background photons in their follow-up study in Ref.~\onlinecite{OE19(2011)}. In Ref.~\onlinecite{OE20(2012)}, the authors studied the cross-correlation between the two paired photons. Their biphoton source had a generation rate of $\sim$$10^7$/s and a temporal FWHM of 1~ns. In Ref.~\onlinecite{Optica2(2015)}, the authors generated photon triplets using the processes of SFWM and spontaneous parametric down-conversion. Their C-band photon pairs had a temporal FWHM of 1~ns. In Ref.~\onlinecite{OE28(2020)}, the authors focused on generating high-dimensional entangled states using the biphotons with a temporal FWHM of 1~ns. In Ref.~\onlinecite{Adv.Q7(2024)}, the authors reported a biphoton source with a generation rate per pump power of 38000/s/mW and a temporal FWHM of 0.6~ns. In Ref.~\onlinecite{PRA21(2024)}, the authors studied the heralding efficiency as a function of the two-photon detuning and optical depth, and their biphoton source had a temporal FWHM of 0.5~ns. So far, all diamond-type or cascade-type biphoton sources using hot atoms have temporal FWHMs equal to or less than 1~ns.

Here, we systematically studied the temporal width of biphotons generated via the diamond-type SFWM process. The biphoton source in this study consisted of a hot atomic medium. When we applied the 795-nm coupling and 1476-nm pump laser fields to the atoms, the SFWM-generated signal and idler photon pairs had wavelengths of 1529-nm and 780-nm, respectively. We developed a theoretical model for the two-photon correlation function, or biphoton wave packet, using the optical Bloch equation of the atomic density-matrix operator. The theoretical predictions of temporal width agree with the experimental data. We further introduced a new concept that enables the hot-atom diamond-type or cascade-type SFWM biphoton source to behave like the cold-atom one. The concept introduces the parameters of the atomic velocity group associated with the two-photon resonance condition and the one-photon detuning in the atom frame. We explain the physical origin of the temporal width as it varies with the parameters. 

\Table

In this work, we have achieved a C-band HSP temporal width of 28.3$\pm$0.6~ns with the atomic velocity group near zero. The biphotons had a peak cross-correlation of 3.00$\pm$0.05, which violates the Cauchy-Schwarz inequality for classical light by a factor of 2.3. Table~\ref{table:One} lists all diamond-type or cascade-type SFWM biphoton sources reported in the literature. The temporal FWHM of our hot-atom biphoton source is the longest in the table and represents the first result exceeding 10~ns. Its corresponding spectral linewidth is 11.0$\pm$0.2~MHz, while the ultimate linewidth, determined by the natural linewidth of the atoms, is 6.1 MHz. By selecting a suitable atomic velocity group, we can significantly improve the biphotons' peak cross-correlation to 9.5$\pm$0.6 while maintaining the HSP temporal FWHM at 17.5$\pm$2.3~ns. This study marks a significant breakthrough in the C-band HSP temporal width of the atom-based biphoton source and provides a better understanding of the diamond-type or cascade-type SFWM biphoton generation.

\section{Theoratical Model}\label{sec:theory}

The biphoton wave packet or cross-correlation function requires information about the cross-susceptibility and self-susceptibilities of the signal and idler photons. These susceptibilities are related to the optical coherences corresponding to the signal and idler transitions. To obtain the optical coherences, we consider the transition scheme shown in Fig.~\ref{fig:Transition}(a) and solve the steady-state optical Bloch equation (OBE) of the atomic density-matrix operator. Details of the OBE can be found in \AppOBE.

The two-photon correlation function of the biphotons, $G^{(2)}(t)$, as a function of the delay time, $t$, of an idler photon triggered or heralded by a signal photon is given by
\begin{eqnarray}
\label{eq:CCF}
	G^{(2)}(t) &=& \left| 
		\int^\infty_{-\infty} d\Delta_s \frac{e^{i\Delta_s t}}{2\pi}
		\bar{\kappa}(\Delta_s) \,
		{\rm sinc}\left[ \bar{\zeta}(\Delta_s) +\bar{\xi}(\Delta_s) \right] \right.
		\nonumber \\
	&&\times
		\left. e^{i\bar{\xi}(\Delta_s)} \right|^2, 
\end{eqnarray}
where the expressions for $\bar{\kappa}(\Delta_s)$, $\bar{\zeta}(\Delta_s)$, and $\bar{\xi}(\Delta_s)$ can be found in Eqs.~(\ref{eq:kappa})-(\ref{eq:xi}), with $\alpha$ representing the optical depth of the medium and $\Gamma$ = 2$\pi$$\times$6~MHz representing the spontaneous decay rate of states $|2\rangle$ and $|3\rangle$. The square of the integrand in Eq.~(\ref{eq:CCF}) represents the biphoton spectrum, with $|\bar{\kappa}|^2$ indicating the FWM efficiency, $|{\rm sinc}(\bar{\zeta}+\bar{\xi})|^2$ reveals the attenuation due to the phase mismatch, and $|{\rm exp}(i\bar{\xi})|^2$ shows the idler's transmittance. Since the gain for the signal photons is negligible, we neglect $|{\rm exp}(i\bar{\zeta})|^2$ in the biphoton spectrum. 

The SFWM process must be initiated by the population in state $|4\rangle$ ($\rho_{44}$); that is, if $\rho_{44}$ is zero, no biphoton generation occurs. It can be expected that $\rho_{44}$ is significant only for atomic velocity groups centered around the velocity, $v_0$, whose corresponding Doppler shift ensures the coupling and pump fields satisfy the two-photon resonance. That is, $(k_c + k_p) v_0 = \Delta_c + \Delta_p$, where we neglect the AC Stark shifts induced by the coupling and pump fields, with $k_c$ and $k_p$ being the wave vectors of the two fields. The solutions to Eqs.~(\ref{eq:OBE_rho11})-(\ref{eq:OBE_rho41}) also confirm the expectation. The velocity group $v_0$ corresponds to the Doppler shift of the coupling transition, $\omega_{D0}$, i.e., $\omega_{D0} \equiv k_c v_0$, where $v_0 > 0$ indicates that the atoms move in the same direction as the light fields. Thus,
\begin{equation}
\label{eq:omega_D0}
	\omega_{D0} = \frac{\Delta_c + \Delta_p}{1 + (k_p/k_c)}.
\end{equation}
Figure~\ref{fig:Transition}(b) shows the transition diagram in the atom frame of $\omega_{D0}$, where the coupling-pump transition satisfies the two-photon resonance with a one-photon detuning, $\Delta_{\rm{atom}}$, as given by 
\begin{equation}
\label{eq:Delta_atom}
	\Delta_{\rm{atom}} \equiv \Delta_c - \omega_{D0}. 
\end{equation}
The parameter $\omega_{D0}$ represents the atomic velocity group that dominates the SFWM process, and the parameter $\Delta_{\rm atom}$ represents the one-photon detuning in this atom frame. These two parameters are the most influential factors on the biphoton temporal width. Thus, we will use $\omega_{D0}$ and $\Delta_{\rm atom}$ instead of $\Delta_c$ and $\Delta_p$ to specify the conditions in the experiment or calculations throughout the article. 

\FigTwo

A representative biphoton wave packet, calculated from Eq.~(\ref{eq:CCF}), is shown in Fig.~\ref{fig:Theory_WP}(a). The inset shows the biphoton spectrum given by the integrand in Eq.~(\ref{eq:CCF}). The wave packet exhibits two decays with significantly different time constants, which we refer to as the slow decay (with a much longer time constant) and the fast decay (with a much shorter time constant). Since this work emphasizes temporally long or spectrally narrow-linewidth biphotons (or C-band HSPs), the fast decay part of the biphoton wave packet will be described elsewhere. We employed etalons in the experiment to filter out the fast decay part and used Eq.~(\ref{eq:GtwoEtalon}) to calculate the biphoton wave packet after the etalons as shown in Fig.~\ref{fig:Theory_WP}(b). The blue line represents the best fit of a phenomenological function given by 
\begin{equation}
\label{eq:fitting}
	A \left[ 1 + \tanh\left( \frac{t-t_0}{\tau_1} \right) \right]^p 
	 	\left[ 1 - {\rm erf}\left( \frac{t-t_0 - t_d}{\tau_2} \right) \right],
\end{equation}
where ${\rm erf}(\,)$ is the error function, and $A$, $t_0$, $p$, $\tau_1$, $\tau_2$, and $t_d$ are the fitting parameters. We obtain the theoretical prediction of the temporal FWHM, $\tau_b$, from the best fit.

\FigThree

We systematically studied $\tau_b$ as a function of $\Delta_{\rm atom}$ at several values of $\omega_{D0}$, as shown in Fig.~\ref{fig:Theory_Delta}. The solid lines reveal that a larger $\Delta_{\rm atom}$ produces a longer $\tau_b$, while a greater $\omega_{D0}$ results in a shorter $\tau_b$. One can understand the behavior from the biphoton spectrum, which is the product of the spectra of the FWM efficiency $|\bar{\kappa}|^2$, the degree of phase mismatch $|{\rm sinc}(\bar{\zeta}+\bar{\xi})|^2$, and the idler's transmission $|{\rm exp}(i\bar{\xi})|^2$. The spectral contrast is defined by (maximum$-$minimum)/(maximum$+$minimum) of a spectrum. The contrast of the spectrum $|\bar{\kappa}|^2$ is always 1, making it a decisive factor in the biphoton spectrum. As $\omega_{D0}$ or $\Delta_{\rm atom}$ increases, the contrast of the spectrum $|{\rm exp}(i\bar{\xi})|^2$ decreases. The spectrum $|{\rm exp}(i\bar{\xi})|^2$ is influential only when both $\omega_{D0}$ and $\Delta_{\rm atom}$ are small. The contrast of the spectrum $|{\rm sinc}(\bar{\zeta}+\bar{\xi})|^2$ is always small, so it is never a decisive factor in the biphoton spectrum. For a fixed $\omega_{D0}$, a larger $\Delta_{\rm atom}$ results in a narrower linewidth of $|\bar{\kappa}|^2$. Consequently, as $\Delta_{\rm atom}$ increases, the biphoton spectral linewidth narrows, and $\tau_b$ becomes longer. For a fixed $\Delta_{\rm atom}$, $\omega_{D0}$ has little effect on the linewidth of $|\bar{\kappa}|^2$, and a smaller $\omega_{D0}$ makes the linewidth of $|{\rm exp}(i\bar{\xi})|^2$ more influential. Consequently, as $\omega_{D0}$ decreases, the biphoton spectrum becomes narrower due to $|{\rm exp}(i\bar{\xi})|^2$, and $\tau_b$ increases. Thus, the spectra of $|\bar{\kappa}|^2$ and $|{\rm exp}(i\bar{\xi})|^2$ explain how $\omega_{D0}$ and $\Delta_{\rm atom}$ affect $\tau_b$.

In Fig.~\ref{fig:Transition}(b), the two-photon transition driven by the coupling and pump fields excites $\rho_{44}$, initiating the SFWM process. Since the one-photon detuning $\Delta_{\rm atom}$ of the two-photon transition is typically large, the excitation exhibits an effective Rabi frequency, $\Omega_{\rm eff}$, given by
\begin{equation}
\label{eq:Omega_eff}
        \Omega_{\rm{eff}} = \frac{\Omega_c \Omega_p}{2 \Delta_{\rm{atom}}},
\end{equation}
where $\Omega_c$ and $\Omega_p$ are the coupling and pump Rabi frequencies. As long as $\Omega_{\rm eff}$ remains the same, one can expect that different combinations of $\Omega_c$ and $\Omega_p$ will produce very similar results. In Fig.~\ref{fig:Theory_Delta}, the central solid line and two nearby dashed lines confirm the expectation. 

Furthermore, a larger $\Omega_{\rm eff}$ resulting from smaller $\Delta_{\rm atom}$ leads to more velocity groups around $\omega_{D0}$ having significant $\rho_{44}$, contributing to the SFWM process. A wider range of velocity groups participating in the SFWM process broadens the linewidth of the FWM spectrum $|\bar{\kappa}|^2$. As $\Delta_{\rm atom}$ increases, $\Omega_{\rm eff}$ decreases, causing the number of velocity groups with significant $\rho_{44}$ to decrease, thereby narrowing the linewidth of $|\bar{\kappa}|^2$. As $\Delta_{\rm atom}$ becomes very large, the natural linewidth of atoms, $\Gamma$, and the decoherence rate, $\gamma$, determine the velocity groups with significant $\rho_{44}$. The linewidth of $|\bar{\kappa}|^2$, or equivalently the biphoton spectrum, asymptotically approaches its minimum. That is, $\tau_b$ reaches its maximum. According to $\Omega_{\rm eff}$, a larger (smaller) value of $\Omega_c$$\times$$\Omega_p$, i.e., the lower (upper) solid line of the same color in Fig.~\ref{fig:Theory_Delta}, has an effect similar to that of a smaller (larger) value of $\Delta_{\rm atom}$.'

\FigFour

We show the effect of optical depth (OD) on $\tau_b$ in Fig.~\ref{fig:Theory_OD}. A higher OD results in a larger value of $\tau_b$, as shown by each set of three lines of the same color. The idler transmission spectrum $|{\rm exp}(i\bar{\xi})|^2$ is responsible for this effect. A higher OD narrows the linewidth of $|{\rm exp}(i\bar{\xi})|^2$, thereby reducing the biphoton linewidth and enhances $\tau_b$. As $\omega_{D0}$ increases, i.e., as the number of atoms interacting with the idler photons decreases, the contrast of $|{\rm exp}(i\bar{\xi})|^2$ becomes smaller, making it less influential on the biphoton spectrum. Consequently, the OD enhancement for narrowing the biphoton spectral linewidth or prolonging the biphoton temporal width becomes less prominent.

The dephasing or decoherence rate, $\gamma$, in the medium of hot atoms is often nonnegligible due to relaxation processes caused by thermal motion. In Fig.~\ref{fig:Theory_gamma}, the blue (or red) lines with $\omega_{D0}/2\pi$ = 0 (or 0.48~GHz) show $\tau_b$ as a function of $\Delta_{\rm{atom}}$ under different values of $\gamma$. As $\gamma$ increases, the linewidth of the biphoton spectrum becomes broadened, shortening the temporal width. When one reduces $\Delta_{\rm{atom}}$, $\Omega_{\rm eff}$ becomes significantly larger than $\gamma$ and dominates the value of $\tau_b$, such that $\gamma$ has little influence on $\tau_b$. Furthermore, the asymptotic value of $\tau_b$ at very large $\Delta_{\rm atom}$ is also reduced by $\gamma$, and is approximately given by $1/(\Gamma + 2\gamma)$. A larger value of $\omega_{D0}$ causes $\tau_b$ to approach the asymptotic value at larger $\Delta_{\rm{atom}}$.

In this section, we theoretically studied the biphoton wave packet's $\tau_b$ (the temporal FWHM) as a function of $\omega_{D0}$ (the atomic velocity group dominating the SFWM process) and $\Delta_{\rm atom}$ (the one-photon detuning in this atom frame). These two parameters play a decisive role in $\tau_b$ and are equivalent to $\Delta_c$ and $\Delta_p$ (the one-photon detunings of the coupling and pump fields) as shown by Eqs.~(\ref{eq:omega_D0}) and (\ref{eq:Delta_atom}). We also investigated the effects of $\Omega_{\rm eff}$ [the effective Rabi frequency defined by Eq.~(\ref{eq:Omega_eff})], $\gamma$ (the decoherence rate), and $\alpha$ (the OD) on $\tau_b$. These studies helped us design the experiment and identify the maximum achievable temporal width of biphotons or C-band HSPs.

\FigFive

\section{Experimental Setup}\label{sec:setup}

We employed a vapor cell of isotopically enriched $^{87}$Rb atoms in the experiment. The temperature of the vapor cell was maintained at 55 $^\circ$C, and its inner wall has a paraffin coating. The $e^{-1}$ half-width of a Doppler-broadened 795-nm transition line for these hot atoms was approximately 0.32~GHz. Figure~\ref{fig:Transition}(a) shows the diamond-type SFWM transition diagram employed in the experiment, where states $|1\rangle$, $|2\rangle$, $|3\rangle$, and $|4\rangle$ represent the energy levels $\ket{5S_{1/2}, F=2}$, $\ket{5P_{1/2}, F=2}$, $\ket{5P_{3/2}, F=3}$, and $\ket{4D_{3/2}, F=3}$ of $^{87}$Rb atoms, respectively. The wavelengths of the coupling and pump fields are 795 and 1476 nm, and those of the signal and idler single photons are 1529 and 780 nm.

The coupling field was generated using a Toptica DL DLC pro laser with the $s$ polarization. Its frequency was stabilized using the saturated absorption spectrum with the Pound-Drever-Hall (PDH) scheme. The pump field was generated using another Toptica DL DLC pro laser with the $p$ polarization. Its power was boosted using a FiberLabs AMP-FL8212-SB-20-PM-NT1 fiber amplifier. The pump frequency was stabilized using the spectrum of the coupling-pump two-photon transition $|1\rangle \rightarrow |2\rangle \rightarrow |4\rangle$ with the PDH scheme. The two-photon transition exhibited a stability of approximately 340~kHz measured at a bandwidth of 100~kHz. The frequency stabilization setup described here is similar to that outlined in Ref.~\onlinecite{OurPRAppl2024}.

The coupling and pump fields, along with the signal and idler photons, all propagated in the same direction, and they completely overlapped in the vapor cell, as depicted in Fig.~\ref{fig:ExpSetup}. This all-copropagation configuration ensured excellent phase-matching and minimized the decoherence rate induced by the Doppler effect \cite{OurOpEx2021, OurQST2025}. The signal and idler photons were polarized in the $s$ and $p$ directions, respectively. After emerging from the vapor cell, the two kinds of photons were separated by a dichroic mirror (Thorlabs DMSP1180) and collected into two polarization-maintained optical fibers. Bandpass filters (Thorlabs FBH780-10 for 780 nm and Thorlabs FBH1530-12 for 1529 nm) and polarization beam splitters were used to block the coupling and pump fields. The leakage rate of the coupling or pump laser ($<30$ or $<10$ counts/s, respectively) into each single-photon counting module (SPCM) was negligible compared to the idler or signal count rates (7$\sim$25 or 2$\sim$17 $\times10^3$ counts/s, respectively). In this work, we maintained the coupling and pump powers at 30 and 60~mW, respectively, close to their maximum available values. 

\FigSix

The signal SPCM (IDQube NIR-FR-MMF-LN) exhibited a quantum efficiency of 12\% and a dead time of 10~$\mu$s. Its dark count rate was about 270 counts/s. The idler SPCM (Excelitas SPCM-AQRH-13-FC) exhibited a quantum efficiency of 49\% and a dead time of 24~ns. Its dark count rate was about 180 counts/s. The outputs of the two SPCMs were recorded using a time tagger (IDQ ID900-MASTER), which generated a histogram of coincidence counts. We used the idler photon counts to trigger the time tagger and employed a long BNC cable between the signal SPCM and the time tagger to introduce a delay in the signal photon counts \cite{OurQST2025}. Each trigger from an idler photon count initiated a 200~ns time window to record the arrival time of a signal photon count.

We employed a 1529-nm etalon and a 780-nm etalon to filter out the fast decay part of the biphoton wave packet, as shown in Fig.~\ref{fig:Theory_WP}. The two etalons also selected the desired transitions of the signal and idler photons. They had a combined spectral FWHM of 38~MHz, which had a minimal effect on the C-band HSP's maximum temporal FWHM of 25.7~ns achieved in this work. The peak transmittance of the 1529-nm etalon was 42\%, and that of the 780-nm etalon was 70\%. \AppEtalon~illustrates the transmission spectra of the two etalons.

The detection efficiencies of the signal and idler photons were 7.0$\pm$0.6\% and 25$\pm$1\%, respectively. These values account for the attenuation caused by optical components in the optical path and the quantum efficiency of the SPCMs. These values exclude the collection efficiency of the two polarization-maintained optical fibers and the transmittances of the two etalons. The signal and idler optical fibers had collection efficiencies of 82\% and 74\% for the spatial modes of the pump and coupling laser beams, respectively. The net transmittance of the two etalons depended on the temporal width of the biphoton wave packet, which will be illustrated in \AppEtalon.

\section{Results and Discussion}\label{sec:results}

\FigSeven

When we measured the biphoton wave packets or two-photon correlation functions, the idler photons acted as heralding photons, triggering the process. The C-band signal photons were the heralded photons and generated coincidence counts upon receiving the triggers. Figures~\ref{fig:Data_WP}(a) and \ref{fig:Data_WP}(b) show the representative data of the biphoton wave packet, i.e., the coincidence counts as a function of the signal photon's delay time, measured at $\omega_{D0}/2\pi$ = 0 and 0.48~GHz, respectively, where $\omega_{D0}$ represents the center Doppler shift of the atomic velocity groups participating in the biphoton generation and its relation to the coupling and pump detunings is given by Eq.~(\ref{eq:omega_D0}). The biphoton temporal profile in each subfigure is time-reversed compared to its counterpart, using the signal photons as the heralding photons, e.g., the biphoton temporal profiles shown in Fig.~\ref{fig:Theory_WP}. The temporal FWHMs, $\tau_b$, of the two kinds of temporal profiles are the same \cite{OurQST2025}. To remove the fast decay peak [see the example shown in Fig.~\ref{fig:Theory_WP}(a)], the 1529-nm and 780-nm etalons were used in all measurements throughout this study.

The absorption spectrum of a weak idler laser beam driving the transition from $|5S_{1/2}, F=2\rangle$ to $|5P_{3/2}, F=3\rangle$ in the presence of the coupling and pump fields determines the OD of the atomic vapor. \AppOD~describes the details of the measurement and how the OD varies with $\Delta_{\rm atom}$, where $\Delta_{\rm atom}$ represents the one-photon detuning in the atom frame that participates in the biphoton generation and its relation to the coupling and pump detunings is described in Eq.~(\ref{eq:Delta_atom}).

We utilized the phenomenological function in Eq.~(\ref{eq:fitting}) to fit the experimental data of the biphoton wave packet. The best fits agree well with the data, as demonstrated by the solid red lines in Figs.~\ref{fig:Data_WP}(a) and \ref{fig:Data_WP}(b). Each inset shows the spectrum of the best fit in the corresponding subfigure. The FWHMs from the best fit and their spectrum provide temporal and spectral widths. The uncertainty specified in the caption reflects the fluctuation observed across multiple measurements under identical experimental conditions. 

In the absence of triggers from the heralding photons, the heralded photons exhibit the nature of thermal light. \AppUACF~describes the Hanbury-Brown-Twiss (HBT) measurement of the auto-correlation function, $g^{(2)}_{s,s}(t)$. The HBT measurement determined $g^{(2)}_{s,s}(0) =$ 2.06$\pm$0.04, revealing that the untriggered or free-run signal photons are thermal light. Similarly, $g^{(2)}_{i,i}(0) =$ 1.95$\pm$0.05, indicating that the untriggered or free-run idler photons also display thermal light behavior. The detection rate of the temporally long biphotons was insufficient to measure the three-fold coincidence count within a reasonable time scale. Hence, we did not measure the conditional auto-correlation function of the signal photons, $g^{(2)}_{s,s|i=1}(t)$. However, as detailed in Ref.~\onlinecite{OurOPEX2024}, one can utilize the cross-correlation function, $g^{(2)}_{s,i}(t)$, and the value of $g^{(2)}_{s,s}(0)$ to derive $g^{(2)}_{s,s|i=1}(t)$.

\FigEight

We systematically measured the biphotons' $\tau_b$ as a function of $\Delta_{\rm atom}$ at two representative values of $\omega_{D0}/2\pi$, namely 0 and 0.48~GHz, where the $e^{-1}$ Doppler half-width was 0.32~GHz. The circles in Fig.~\ref{fig:Data_tFWHM} are the experimental data, and the solid lines are the theoretical predictions. We varied the values of $\Omega_c$$\times$$\Omega_p$ and $\gamma$ to match the predictions only to the data with $\Delta_{\rm atom}/2\pi >$ 0.14~GHz. The ODs of these data points were nearly the same. Therefore, the variation in OD had minimal impact on the determination of $\Omega_c$$\times$$\Omega_p$ and $\gamma$. We extended the theoretical predictions to the data points with $\Delta_{\rm atom}/2\pi <$ 0.14~GHz. All the ODs in the calculations of the theoretical predictions were determined experimentally.  

The general trends of the theoretical predictions are consistent with those of the experimental data in Fig.~\ref{fig:Data_tFWHM}. These trends include a larger $\Delta_{\rm atom}$ resulting in a longer $\tau_b$, a larger $\omega_{D0}/2\pi$ leading to a shorter $\tau_b$, $\tau_b$ having an upper limit at very large $\Delta_{\rm atom}$, and $\tau_b$ asymptotically approaching the limit at  $\omega_{D0}$ = 0 faster than $\omega_{D0}/2\pi$ = 0.48~GHz. The shaded areas with the $\pm$10\% changes in $\alpha$ and $\Omega_c$$\times$$\Omega_p$ shown in the figure account for the uncertainties in OD and Rabi frequencies. The theoretical model assumes plane-wave propagation. Although we kept the same coupling and pump powers during all the measurements, their effective Rabi frequencies in the plane-wave model may vary, introducing uncertainty. The determination of the OD, or equivalently atomic density, relies on the fitting of an absorption spectrum contributed by the entire atomic ensemble. Hence, the determined atomic density might not exactly correspond to the atoms in the velocity group participating in the SFWM process, which introduces additional uncertainty. The quantitative consistency between the theoretical predictions (with the shaded areas) and the experimental data is satisfactory.

\FigNine

The same data for biphoton wave packets, i.e., the coincidence count versus the delay time, which determined $\tau_b$ in Fig.~\ref{fig:Data_tFWHM}, also provided the biphoton detection rates. To derive the generation rate from the detection rate, we took into account the SPCMs' quantum efficiencies and dead times, the etalons' attenuation (see \AppEtalon~for the correction method), and the loss of all optical components. The derivation did not account for the collection efficiencies of the two optical fibers that collect the idler and signal photons emitted from the atomic vapor cell. In Fig.~\ref{fig:Data_GR}, the generation rate refers to the biphotons collected by polarization-maintained optical fibers. The circles in the figure are the experimental data of the biphoton generation rate versus $\Delta_{\rm atom}$, with the solid lines showing the theoretical predictions. Using Eq.~(\ref{eq:Gtwo}), we calculated the theoretical predictions here with the same parameters of $\alpha$, $\Omega_c$, $\Omega_p$, and $\gamma$ as those in Fig.~\ref{fig:Data_tFWHM}. The result of Eq.~(\ref{eq:Gtwo}) is proportional to the value of the solid line in Fig.~\ref{fig:Data_GR}. We determined the proportionality by matching the theoretical predictions to the data with $\Delta_{\rm atom}/2\pi >$ 0.14~GHz. The shaded areas in Fig.~\ref{fig:Data_GR} correspond to $\pm$10\% changes in $\alpha$ and $\Omega_c$$\times$$\Omega_p$.

The deviation of the data points at small values of $\Delta_{\rm atom}$ from the blue and red lines in Fig.~\ref{fig:Data_GR} may be due to the uncertainties in the OD and Rabi frequencies. These uncertainties were previously mentioned to explain the shaded areas in Fig.~\ref{fig:Data_tFWHM}. Furthermore, the proportionality discussed earlier for the red line is 1.9 times larger than that for the blue line. That is, if the generation rate data for $\omega_{D0}$ = 0 are consistent with the theoretical predictions, then those for $\omega_{D0}/2\pi$ = 0.48~GHz are 1.9 times higher than expected. Since this work focused on the biphoton temporal width, the deviations in the generation rate between the data and predictions will be discussed in future work.  

\FigTen

We plot the peak of the cross-correlation function between the signal and idler photons, $g^{(2)}_{s,i}(t_0)$, as a function of $\Delta_{\rm atom}$ in Fig.~\ref{fig:Data_CCF}. The same biphoton wave packets that determined $\tau_b$ in Fig.~\ref{fig:Data_tFWHM} and the generation rate in Fig.~\ref{fig:Data_GR} also provide $g^{(2)}_{s,i}(t_0)$ here. Figure~\ref{fig:Data_CCF} reveals two phenomena. First, $g^{(2)}_{s,i}(t_0)$ decreases with $\Delta_{\rm atom}$ at given $\omega_{D0}$. Second, $g^{(2)}_{s,i}(t_0)$ at $\omega_{D0}/2\pi =$ 0.48~GHz is significantly larger than that at $\omega_{D0} =$ 0. The first phenomenon, together with the result of the generation rate decreasing with $\Delta_{\rm atom}$ as shown in Fig.~\ref{fig:Data_GR}, indicates that the value of $g^{(2)}_{s,i}(t_0)$ increases with the generation rate at given $\omega_{D0}$. We infer that the noise or background photons mainly arise from two sources: the unpaired photons with their counterparts being absorbed by atoms and the spontaneously emitted photons without the FWM process. The background photon rate did not vary much against $\Delta_{\rm atom}$ (the variation affected the background level by $\leq\pm$44\% or $\leq\pm$27\% at  $\omega_{D0}/2\pi$ = 0 or 0.48~GHz) compared to the increase in the generation rate, which explains why a higher generation rate results in a larger value of $g^{(2)}_{s,i}(t_0)$. The second phenomenon may be due to far less atomic density at $\omega_{D0}/2\pi =$ 0.48~GHz. Thus, the background photon rate dropped more significantly than the generation rate, which explains that $g^{(2)}_{s,i}(t_0)$ at $\omega_{D0}/2\pi =$ 0.48~GHz was significantly higher.

\FigEleven

To minimize the decoherence rate, we utilized the classical-light FWM spectrum to optimize the alignments of the coupling and pump fields' propagation directions and the signal and idler photons' collection directions. In measuring the FWM spectrum, a 1529-nm laser field was directed at the atoms in the presence of the coupling and pump fields, and a 780-nm field was generated at the output. The collection directions of the signal and idler photons  were aligned to the propagation directions of the 1529-nm and 780-nm fields. We adjusted the alignments to maximize the detected 780-nm power. Figure~\ref{fig:LongestWP} shows the biphoton wave packet measured after the optimization. The temporal width became more prolonged compared to before the optimization under the same experimental parameters, revealing that the optimization reduces the decoherence rate. As shown in the figure, the theoretical prediction, which best matches the data, also confirms the reduction of the decoherence rate. As $\Delta_{\rm atom}$ further increased, the biphoton temporal width became slightly longer, but the peak cross-correlation decreased. Thus, the temporal width at $\Delta_{\rm atom}/2\pi =$ 2.12~GHz is now very close to the maximum value under the present decoherence rate. Because of the alignment optimization, we also improved the biphoton generation rate by 2.2 times. The decoherence rate reduction cannot explain this improvement. It indicates that the optimization diminished the phase mismatch induced by the misalignment, which is not considered in the theoretical model. In this work, we have achieved a biphoton source of C-band heralded single photons with a temporal FWHM of about 28~ns, corresponding to a linewidth of 11~MHz, at a generation rate of 7.6$\times$$10^3$ pairs/s.

\section{Conclusion}

Since the realization of the first diamond-type or cascade-type SFWM biphoton source\cite{PRL96(2006)}, to our knowledge, no paper has comprehensively studied the theoretical model of biphoton generation in either Doppler-free or Doppler-broadened media. We systematically studied the theory of diamond-type SFWM biphoton generation in a room-temperature or hot atomic medium. The developed theoretical framework can also be applied to laser-cooled or cold atomic media, as well as to the cascade-type transition scheme with minimal modification.

Our theoretical and experimental studies focused on the biphoton temporal width. We introduced the two important parameters: $\omega_{D0}$ (the atomic velocity group dominating the SFWM process) and $\Delta_{\rm atom}$ (the one-photon detuning in this atom frame). These parameters play a decisive role in the biphoton temporal FWHM, $\tau_b$; they are equivalent to $\Delta_c$ and $\Delta_p$ (the one-photon detuning of the coupling and pump fields) and are more intuitive than $\Delta_c$ and $\Delta_p$ for understanding the underlying mechanism. The experimental data of $\tau_b$ as a function of $\Delta_{\rm atom}$ at two representative values of $\omega_{D0}$ are consistent with the theoretical predictions based on the developed theoretical framework.

With the guidance of the theory, we achieved a temporal FWHM of C-band heralded single photons at 28.3~ns, marking the first result to exceed 10 ns among all diamond-type or cascade-type SFWM biphoton sources utilizing either cold or hot atoms. The corresponding linewidth is 11~MHz, where the natural linewidth of the atoms sets the ultimate limit of the biphoton linewidth to 6 MHz. Since atom-based biphoton sources have the advantages of narrow linewidth, stable frequency, and highly tunable bandwidth, this work provides a better understanding of the diamond-type or cascade-type SFWM biphoton generation. It advances the technology of long-distance quantum communication utilizing the C-band heralded single photons or qubits.

\section*{ACKNOWLEDGMENTS}
This work was supported by Grants No.~112-2112-M-007-020-MY3 and No.~113-2119-M-007-012 of the National Science and Technology Council, Taiwan.

\section*{CONFLICT OF INTEREST}
The authors declare no conflicts of interest.

\section*{DATA AVAILABILITY}
The data that support the findings of this study are available from the corresponding author upon reasonable request.

\appendix

\section{Predictions of Biphoton Wave Packet and the Optical Bloch Equation} \label{sec:OBE}

For the diamond-type SFWM transition scheme shown in Fig.~\ref{fig:Transition}(a), the steady-state optical Bloch equations (OBEs) of the density-matrix operator of atoms are given by
\begin{eqnarray}
\label{eq:OBE_rho11}
	1 &=& 
	\rho_{11} +\rho_{22} +\rho_{33} +\rho_{44}, 
	\\
\label{eq:OBE_rho22}
	0 &=& 
	-\Gamma_{2}\rho_{22} +\frac{\Gamma_4}{2}\rho_{44}
	\nonumber \\ && 
	+\frac{i}{2}( \Omega_c\rho_{21}^\ast -\Omega_c^\ast\rho_{21}
		+\Omega_p^\ast\rho_{42} -\Omega_p\rho_{42}^\ast ), 
	\\
\label{eq:OBE_rho33}
	0 &=& 
	-\Gamma_{3}\rho_{33} +\frac{\Gamma_4}{2}\rho_{44}, 
	\\
\label{eq:OBE_rho44}
	0 &=& 
	-\Gamma_4\rho_{44} 
	+\frac{i}{2}( \Omega_p\rho_{42}^\ast -\Omega_p^\ast\rho_{42} ), 
	\\
\label{eq:OBE_rho21}
	0 &=& 
	i\left( \Delta_c-\omega_D \right) \rho_{21}
	-\left( \frac{\Gamma_2}{2}+\gamma \right) \rho_{21} 
	\nonumber \\ &&
	+\frac{i}{2}( \Omega_c\rho_{11} -\Omega_c \rho_{22} 
		+\Omega_p^\ast\rho_{41} ), 
	\\
\label{eq:OBE_rho42}
	0 &=& 
	i\left( \Delta_p-\frac{k_p}{k_c}\omega_{D} \right) \rho_{42}
	-\left(\frac{\Gamma_{2}}{2} +\frac{\Gamma_{4}}{2} +\gamma\right)\rho_{42} 
	\nonumber \\ &&	
	+\frac{i}{2}( \Omega_p\rho_{22} -\Omega_p\rho_{44}
		-\Omega_c^\ast\rho_{41} ), 
	\\
\label{eq:OBE_rho41}
	0 &=& 
	i\left( \Delta_c -\omega_D
		+\Delta_p-\frac{k_p}{k_c}\omega_D \right) \rho_{41}
	-\left( \frac{\Gamma_4}{2} +\gamma \right) \rho_{41}
	\nonumber \\ &&	
	+\frac{i}{2}( \Omega_p\rho_{21} -\Omega_c\rho_{42}), 
\end{eqnarray}
\begin{eqnarray}
\label{eq:OBE_rho43}
	0 &=& 
	i\left( \Delta_s-\frac{k_s}{k_c}\omega_D \right) \rho_{43}
	-\left(\frac{\Gamma_3}{2}+\frac{\Gamma_4}{2} +\gamma\right)\rho_{43}
	\nonumber \\ &&	
	+\frac{i}{2} (\Omega_s\rho_{33} -\Omega_s\rho_{44} +\Omega_p\rho_{32}^\ast
		-\Omega_i^\ast\rho_{41} ),  
	\\
\label{eq:OBE_rho32}
	0 &=& 
	i\left( \Delta_p -\frac{k_p}{k_c}\omega_D -\Delta_s
		+\frac{k_s}{k_c}\omega_D \right) \rho_{32}
	\nonumber \\ &&	
	-\left( \frac{\Gamma_3}{2} +\frac{\Gamma_2}{2} +\gamma \right) \rho_{32} 
	\nonumber \\ &&	
	+\frac{i}{2}( \Omega_s^\ast\rho_{42} +\Omega_i\rho_{21}^\ast
		-\Omega_c^\ast\rho_{31} -\Omega_p\rho_{43}^\ast ), 
	\\
\label{eq:OBE_rho31}
	0 &=& 
	i\left( \Delta_c-\omega_D +\Delta_p -\frac{k_p}{k_c}\omega_D 
		-\Delta_s +\frac{k_s}{k_c}\omega_D \right) \rho_{31} 
	\nonumber \\ &&	
	-\left( \frac{\Gamma_3}{2}+\gamma \right) \rho_{31} 
	\nonumber \\ &&	
	+\frac{i}{2}( \Omega_i\rho_{11} -\Omega_i\rho_{33} -\Omega_c\rho_{32}
		+\Omega_s^\ast\rho_{41} ),
\end{eqnarray}
where $\rho_{ij}$ ($i,j$ = 1, 2, 3, or 4) is an element of the density-matrix operator, $\Gamma_2$, $\Gamma_3$, and $\Gamma_4$ are the spontaneous decay rates of states $|2\rangle$, $|3\rangle$, and $|4\rangle$, $\gamma$ denotes the dephasing or decoherence rate of optical coherences, $\Omega_c$, $\Omega_p$, $\Omega_s$, and $\Omega_i$ are the coupling, pump, signal, and idler Rabi frequencies, $\Delta_c$, $\Delta_p$, and $\Delta_s$ are the coupling, pump, and signal detunings, $\omega_D$ represents the Doppler shift, and $k_c$, $k_p$, and $k_s$ denote the wave vectors for the coupling, pump, and signal, respectively.

To derive the OBEs, we consider the signal and idler photons as perturbations. Equations~(\ref{eq:OBE_rho11})-(\ref{eq:OBE_rho41}) represent the zeroth-order OBEs, and Eqs.~(\ref{eq:OBE_rho43})-(\ref{eq:OBE_rho31}) are the first-order ones. Note that the idler detuning, $\Delta_i$, can be derived from the other three detunings due to four-photon resonance, i.e., the term $\Delta_c-\omega_D +\Delta_p -(k_p/k_c)\omega_D  -\Delta_s +(k_s/k_c)\omega_D$ in Eq.~(\ref{eq:OBE_rho31}) is equal to the idler detuning in the atomic frame. 

In Eqs.~(\ref{eq:OBE_rho11})-(\ref{eq:OBE_rho44}), we assume that the population in $|4\rangle$ decays equally to $|2\rangle$ and $|3\rangle$, and the populations of $|2\rangle$ and $|3\rangle$ decay only to $|1\rangle$. The assumption affects the quantitative results but does not alter the underlying physics of the biphoton generation. We set $\Gamma_2$ and $\Gamma_3$ to $\Gamma$ and made $\Gamma_4$ equal to $\Gamma/3$ in the calculation, where $\Gamma$ = 2$\pi$$\times$6~MHz. The values of $\Gamma_2$, $\Gamma_3$, and $\Gamma_4$ approximate the spontaneous decay rates of the actual energy levels used in the experiment.

The optical coherences $\rho_{43}$ and $\rho_{31}$ correspond to the transitions of the signal and idler photons, respectively. Given the Rabi frequencies and detunings of the coupling and pump laser fields, the solutions for $\rho_{43}$ and $\rho_{31}$ from the OBEs are linear functions of the variables $\Omega_s$, $\Omega_s^\ast$, $\Omega_i$, and $\Omega_i^\ast$ due to their perturbative nature. The values of $\omega_D$ and $\Delta_s$ serve as parameters in the functions. We take $\partial \rho_{31}/\partial\Omega_s^\ast$, i.e., the proportionality of $\Omega_s^\ast$ in the linear function, to obtain the cross-susceptibility of the idler photons induced by the signal photons. The self-susceptibilities of signal and idler photons are given by $\partial \rho_{43}/\partial\Omega_s$ and $\partial \rho_{31}/\partial\Omega_i$, respectively. These susceptibilities depend solely on $\omega_D$ and $\Delta_s$, providing information on the FWM efficiency, the degree of the phase mismatch, and the attenuation or gain coefficient of the photons propagating in the medium.

The two-photon correlation function of the biphotons, $G^{(2)}(t)$, as a function of the delay time, $t$, of an idler photon upon the heralding or trigger of a signal photon is given by
\begin{eqnarray}
\label{eq:Gtwo}
	G^{(2)}(t) &=& \left| 
		\int^\infty_{-\infty} d\Delta_s \frac{e^{i\Delta_s t}}{2\pi}
		\bar{\kappa}(\Delta_s) \,
		{\rm sinc}\left[ \bar{\zeta}(\Delta_s) +\bar{\xi}(\Delta_s) \right] \right.
		\nonumber \\
	&&\times
		\left. e^{i\bar{\xi}(\Delta_s)} \right|^2, 
	\\
\label{eq:kappa}
	\bar{\kappa}(\Delta_s) &=& 
		\frac{\alpha\Gamma}{4} \int^\infty_{-\infty} d\omega_D
		\frac{e^{-\omega^2_D/\Gamma^2_D}}{\sqrt{\pi}\Gamma_D} 
		\left[ \frac{\partial \rho_{31}}{\partial\Omega_s^\ast} \right], 
	\\
\label{eq:zeta}
	\bar{\zeta}(\Delta_s) &=& 
		\frac{\alpha\Gamma}{4} \int^\infty_{-\infty} d\omega_D
		\frac{e^{-\omega^2_D/\Gamma^2_D}}{\sqrt{\pi}\Gamma_D} 
		\left[ \frac{\partial \rho_{43}}{\partial \Omega_s} \right],  
\end{eqnarray}
\begin{eqnarray}
\label{eq:xi}
	\bar{\xi}(\Delta_s) &=&
		\frac{\alpha\Gamma}{4} \int^\infty_{-\infty} d\omega_D
		\frac{e^{-\omega^2_D/\Gamma^2_D}}{\sqrt{\pi}\Gamma_D} 
		\left[ \frac{\partial \rho_{31}}{\partial \Omega_i} \right].
\end{eqnarray}
where $\alpha$ is the optical depth of the medium and $\Gamma_D$ is the Doppler width of the ensemble. We average $\partial \rho_{31}/\partial\Omega_s^\ast$, $\partial \rho_{43}/\partial\Omega_s$, and $\partial \rho_{31}/\partial\Omega_i$ over $\omega_D$ to obtain $\bar{\kappa}$, $\bar{\xi}$, and $\bar{\zeta}$, which are functions of $\Delta_s$. The square of the integrand in Eq.~(\ref{eq:CCF}) represents the biphoton spectrum, where $|\bar{\kappa}|^2$ indicates the FWM efficiency, $|{\rm sinc}(\bar{\zeta}+\bar{\xi})|^2$ reveals attenuation due to the phase mismatch, and $|{\rm exp}(i\bar{\xi})|^2$ shows the idler's transmittance. Since the gain for the signal photons is negligible, we neglect $|{\rm exp}(i\bar{\zeta})|^2$ in the biphoton spectrum.

When employing the 1529-nm and 780-nm etalons in the experiment, we revise the formula of $G^{(2)}(t)$ to
\begin{equation}
\label{eq:GtwoEtalon}
	G^{(2)}_e(t) = \left| 
		\int^\infty_{-\infty} d\Delta_s \frac{e^{i\Delta_s t}}{2\pi}
		F(\Delta_s) \frac{\sqrt{T_p}}{1+4\Delta_s^2/\Gamma_e^2} 
	\right|^2,
\end{equation}
where
\begin{equation}
\label{eq:BiphotonSpectrum}
	F(\Delta_s) \equiv 
		\bar{\kappa}(\Delta_s)\,
		{\rm sinc}\left[ \bar{\zeta}(\Delta_s) +\bar{\xi}(\Delta_s) \right] 
		e^{i\bar{\xi}(\Delta_s)},
\end{equation}
$T_p$ is the peak power transmission, and $\Gamma_e$ represents the spectral FWHM of the electric-field transmission coefficient. The measured spectra of the two etalons in \AppEtalon~reveal that the value of $T_p$ is 29\%, and that of $\Gamma_e$ is 2$\pi$$\times$59~MHz.

We first solved the OBEs of Eqs.~(\ref{eq:OBE_rho11})-(\ref{eq:OBE_rho31}) and then performed the integrations in Eq.~(\ref{eq:kappa})-(\ref{eq:xi}) to obtain $\bar{\kappa}$, $\bar{\zeta}$, and $\bar{\xi}$. Finally, we used Eq.~(\ref{eq:Gtwo}) to calculate the theoretical prediction in  Fig.~\ref{fig:Theory_WP}(a) and Eq.~(\ref{eq:GtwoEtalon}) to calculate those in Figs.~\ref{fig:Theory_WP}(b), \ref{fig:Theory_Delta}, \ref{fig:Theory_OD}, \ref{fig:Theory_gamma}, \ref{fig:Data_tFWHM}, and \ref{fig:Data_GR}.

\section{Linewidths of the Etalons} \label{sec:Etalon}

We employed a 1529-nm etalon and a 780-nm etalon to select the signal and idler photons for the desired transitions and to remove the fast decay part of the biphoton wave packet as shown by the theoretical prediction of Fig.~\ref{fig:Theory_WP}(a). Figures~\ref{fig:Data_WP}(a) and \ref{fig:Data_WP}(b) demonstrate the representative experimental data of the biphoton wave packets after passing through the etalons. The experimental data do not show a fast decay part or a sharp falling or rising edge due to the two etalons. 

The circles in Figs.~\ref{fig:Etalon}(a) and \ref{fig:Etalon}(b) show the transmission spectra of the 1529-nm and 780-nm etalons measured with 1529-nm and 780-nm laser fields, respectively. We fitted the spectral data with Lorentzian functions. The red and blue lines in the figures represent the best fits; these fits are in good agreement with the data. We multiplied the red and blue lines to obtain the magenta line in Fig.~\ref{fig:Etalon}(c), representing the combined transmission spectrum of the two etalons. The black line is the best fit for the magenta line. It is a squared Lorentzian function given by
\begin{equation}
\label{eq:SqLorentzian}
	f_e(x) = \left( \frac{\sqrt{T_p}}{1+4x^2/\Gamma_e^2} \right)^2,
\end{equation} 
where $T_p$ is the peak power transmission and $\Gamma_e$ represents the linewidth of the transmitted electric field of the combined spectrum of the two etalons. The FWHM of the squared Lorentzian function of 2$\pi$$\times$38~MHz divided by $\sqrt{\sqrt{2}-1}$ gives $\Gamma_e =$ 2$\pi$$\times$59~MHz.

\FigTwelve

The linewidth of the combined spectrum of the two etalons attenuates the biphoton wave packet, and the attenuation coefficient depends on the biphoton temporal width. To calculate the biphoton generation rate, we derived the attenuation coefficient, $A_e$, caused by the etalons as follows. According to Eq.~(\ref{eq:GtwoEtalon}), the energy of the measured biphoton wave packet after passing through the etalons is proportional to
\begin{equation}
	\int_{-\infty}^{\infty} dt\, G_e^{(2)}(t) = \frac{1}{2\pi}
		\int_{-\infty}^{\infty} d\Delta_s \left| F_e(\Delta_s) \right|^2,
\end{equation}
where
\begin{equation}
	F_e(\Delta_s) = F(\Delta_s) \sqrt{f_e(\Delta_s)},
\end{equation}
and the expressions for $F(\Delta_s)$ and $f_e(\Delta_s)$ are shown in Eqs.~(\ref{eq:BiphotonSpectrum}) and (\ref{eq:SqLorentzian}), respectively. Furthermore, according to Eq.~(\ref{eq:Gtwo}), the energy of the biphoton wave packet before passing through the etalons is proportional to
\begin{equation}
	\int_{-\infty}^{\infty} dt\, G^{(2)}(t) =
		\frac{1}{2\pi} \int_{-\infty}^{\infty} d\Delta_s 
		\frac{\left|  F_e(\Delta_s)\right|^2}{f_e(\Delta_s)}.
\end{equation}
We obtained the expression for $\left| F_e(\Delta_s)\right|^2$ from the Fourier transform of the biphoton wave packet as shown in the inset of Fig.~\ref{fig:Data_WP}(a) or \ref{fig:Data_WP}(b). Finally, the ratio of $\int_{-\infty}^{\infty} dt\, G_e^{(2)}(t)$ to $\int_{-\infty}^{\infty} dt\, G^{(2)}(t)$ gives
\begin{equation}
	A_e = \frac
		{\int_{-\infty}^{\infty} d\Delta_s \left|  F_e(\Delta_s) \right|^2}
		{\int_{-\infty}^{\infty} d\Delta_s \left|  F_e(\Delta_s) \right|^2 / f_e(\Delta_s)}.
\end{equation}
We used the above value of $A_e$ for the combined energy transmittance of the two etalons to derive the experimental data for the biphoton generation rate shown in Fig.~\ref{fig:Data_GR}.

\section{Optical Depth of the Atoms} \label{sec:OD}

The four states $\ket{5S_{1/2}, F=2}$, $\ket{5P_{1/2}, F=2}$, $\ket{5P_{3/2}, F=3}$, and $\ket{4D_{3/2},F=3}$ in the experiment of the diamond-type SFWM transition scheme shown in Fig.~\ref{fig:Transition} do not form a closed system. In addition to $\ket{4D_{3/2},F=3}$ in the system, the coupling and pump fields also excite populations from the ground state $\ket{5S_{1/2}, F=2}$ to the excited states $\ket{4D_{3/2},F=1,2}$ due to the Doppler effect. Consequently, populations in $\ket{4D_{3/2},F=1, 2,3}$ can decay to the other ground state $\ket{5S_{1/2}, F=1}$, and atoms with population in $\ket{5S_{1/2}, F=1}$ are lost from the system. On the other hand, atoms with population in $\ket{5S_{1/2}, F=2}$ flow into the interaction region of the coupling and pump fields and participate in the SFWM process. Thus, the steady-state optical density (OD) in biphoton generation is the balance between the laser fields' excitation and the atomic flow. 

\FigThirteen

The coupling and pump detunings affected the optical pumping, thus influencing the steady-state OD. We measured the OD using the absorption spectrum of weak laser light, which drives the transition from $\ket{5S_{1/2}, F=2}$ to $\ket{5P_{3/2}, F=1,2,3}$. Since the absorption spectrum is sensitive to the atomic densities of the velocity groups around zero but not those with higher velocities due to the Maxwell-Boltzmann distribution, we set the coupling and pump detunings such that the fields interacted with atoms having velocities close to zero. In Fig.~\ref{fig:OD}, the gray circles represent the experimental data of OD as a function of $\Delta_{\rm atom}$. We fitted the data with a phenomenological hyperbolic tangent function. The ODs determined by the best fit were used for the theoretical predictions in Figs.~\ref{fig:Data_tFWHM} and \ref{fig:Data_GR}.

\section{The Auto-Correlation Function of 1529-nm Signal Photons} \label{sec:UACF}

\FigFourteen

We measured the auto-correlation function of the signal photons generated from the diamond-type SFWM biphoton source. After the signal photons passed through the 1529-nm etalon, they were collected into an optical fiber-based 50/50 single-mode beam splitter (Thorlabs TW1550R5A2). The photons were then detected at the two output ports of the beam splitter using two SPCMs of the same model (IDQube NIR-FR-MMF-LN) with identical settings. In this Hanbury-Brown-Twiss (HBT) measurement, the two-fold coincidence count as a function of the delay time between the two photons from the two output ports provides the auto-correlation function.

Figure~\ref{fig:UACF}(a) shows the auto-correlation function of the signal photons generated under the same experimental condition as Fig.~\ref{fig:Data_WP}(a). The time bin width for the measurement was 2.0~ns, and the coincidence counts were accumulated for one hour. We fitted the experimental data with a double exponential decay function. The peak value of the best fit, i.e., $g_{s,s}^{(2)}(0)$, of 2.06$\pm$0.04 indicates that the signal photons emerging from the biphoton source without the triggers from their counterparts exhibited thermal light behavior. Using the same method, we measured the auto-correlation function of the idler photons generated from the biphoton source. The value of $g_{i,i}^{(2)}(0)$ of 1.95$\pm$0.05 confirms that the idler photons similarly exhibited thermal light characteristics.

The two peaks separated by about 105~ns in Fig.~\ref{fig:UACF}(a) are unexpected. To find the origin of the two peaks, we sent a weak laser beam through the 1529-nm etalon and beam splitter and observed the two peaks as shown in Fig.~\ref{fig:UACF}(b). The time bin width in the measurement was 2.0~ns, and the coincidence counts were accumulated for 10 minutes. Using a weaker laser beam increased the prominence of the peak height. The separation of the two peaks was the same and depended neither on the nature of the light source nor on the light power. The two peaks disappeared when the laser beam passed through only the beam splitter. We conclude that the 1529-nm etalon produced the two unexpected peaks. Thus, the fitting for the experimental data in Fig.~\ref{fig:UACF}(a) does not include the data points of the two peaks. 


\end{document}